\documentclass[11pt]{iopart}

\usepackage{iopams}  

\usepackage[latin1]{inputenc}
\usepackage{graphicx}
\usepackage{mathrsfs}
\usepackage{color}
\usepackage{doi}
\usepackage{hyperref}
\usepackage[nice]{nicefrac}
\usepackage{float}
\usepackage{bm}

\def\br{\mathbf{r}}
\def\bk{\mathbf{k}}

\def\be{\begin{eqnarray}}
\def\ee{\end{eqnarray}}

\def\cvec#1{\ensuremath{\bm{\vec #1}}} % custom notation for constant vectors
\def\unit#1{\bm{\hat #1}}

\begin{document}

\title[Synthetic Gauge Fields for Lattices with Multi-Orbital Unit Cells]{Synthetic Gauge Fields for Lattices with Multi-Orbital Unit Cells: Routes towards a $\pi$-flux Dice Lattice with Flat Bands}

\author{Gunnar M\"oller$^{1,\dagger}$ and Nigel R. Cooper$^2$}
\address{$^1$Functional Materials Group, School of Physical Sciences, University of Kent,  Canterbury CT2 7NZ, United Kingdom\\
$^2$TCM Group, Cavendish Laboratory, University of Cambridge, Cambridge CB3 0HE, United Kingdom}

\ead{$^\dagger$G.Moller@kent.ac.uk}
\vspace{10pt}
\begin{indented}
\item[]26 April 2018
\end{indented}

\pacs{
03.75.Lm, %Tunneling, Josephson effect, Bose-Einstein condensates in periodic potentials, solitons, vortices, and topological excitations - important to use this as the primary PACS
67.85.-d, %Ultracold gases, trapped gases (see also 03.75.-b Matter waves in quantum mechanics)
67.85.Hj %Bose-Einstein condensates in optical potentials 
}

\begin{abstract}
We propose a general strategy for generating synthetic magnetic fields in complex lattices with non-trivial connectivity based on light-matter coupling in cold atomic gases. Our approach starts from an underlying optical flux lattice in which a synthetic magnetic field is generated by coupling several internal states. Starting from a high symmetry optical flux lattice, we superpose a scalar potential with a super- or sublattice period in order to eliminate links between the original lattice sites. As an alternative to changing connectivity, the approach can also be used to create or remove lattice sites from the underlying parent lattice. To demonstrate our concept, we consider the dice lattice geometry as an explicit example, and construct a dice lattice with a flux density of half a flux quantum per plaquette, providing a pathway to flat bands with a large band gap. While the intuition for our proposal stems from the analysis of deep optical lattices, we demonstrate that the approach is robust even for shallow optical flux lattices far from the tight-binding limit. 
We also provide an alternative experimental proposal to realise a synthetic gauge field in a fully frustrated dice lattice based on laser-induced hoppings along individual bonds of the lattice, again involving a superlattice potential. In this approach, atoms with a long-lived excited state are trapped using an `anti-magic' wavelength of light, allowing the desired complex hopping elements to be induced in a specific laser coupling scheme for the dice lattice geometry.
We conclude by comparing the complexity of these alternative approaches, and advocate that complex optical flux lattices provide the more elegant and easily generalisable strategy.
\end{abstract}

\maketitle

\tableofcontents

\section{Introduction}
\label{sec:introduction}

The creation of synthetic gauge fields in cold atomic gases provides new opportunities for realising exotic emergent quantum phases \cite{Cooper:2008hx, Cooper:2013jg, Goldman:2014bvb, Goldman:2016fa, Gross:2017do}.
Prominent target phases include vortex lattices \cite{AboShaeer:2001go} and, at high flux density, bosonic counterparts of the continuum fractional quantum Hall states \cite{Cooper:1999bz,Cooper:2001gy}. When both a (synthetic) field and a lattice potential are present, the continuum quantum Hall states are predicted to persist for appreciable flux densities $n_\phi$ per plaquette \cite{Sorensen:2005bt}. In addition, new classes of quantum Hall states, stabilized only due to the presence of a periodic lattice potential, emerge at larger values of $n_\phi$ owing to the underlying structure the Hofstadter spectrum \cite{Kol:1993wv, Palmer:2006km, 2009PhRvL.103j5303M, 2012PhRvL.108y6809H, Moller:2015kg}, and in particular owing to the presence of single-particle bands with higher Chern numbers $|C|>1$ \cite{Moller:2015kg, Andrews:2018il}.

Early experiments on synthetic gauge fields relied on using rotation to emulate magnetic fields \cite{Cooper:2008hx,Fetter:2009fh}. However, in this approach it is exceedingly difficult experimentally to avoid heating due to asymmetric trapping potentials, so the strongly interacting regime of low density in the lowest Landau level remains out of reach. Prompted in part by the exciting outlook for the creation of new phases of matter, there has been much progress with new theoretical proposals and the experimental realizations for schemes of simulating artificial gauge fields \cite{Jaksch:2003ud, Mueller:2004hc, Eckardt:2005bq, Lin:2009us, Gerbier:2010ho, Dalibard:2011gg, Cooper:2011iv}. Further impetus for synthetic fields stems from the prospect of realising topological flat bands in condensed matter systems -- where spin-orbit coupling may provide suitable complex hopping elements in a tight-binding representation -- sharpening the focus on the underlying commonality of flat single particle bands with non-zero Chern number \cite{Thouless:1982kq, Haldane:1988gh,Kane:2005hl,Tang:2011by,Neupert:2011db,Sun:2011dk,Regnault:2011bu}, and more detailed characteristics of their band geometry \cite{Parameswaran:2012cu,Goerbig:2012cz, 2014PhRvB..90p5139R,Jackson:2015fv}.
Currently no clear target systems realising synthetic magnetic flux have been identified in the solid state, while cold atoms provide a range of successful realizations.\footnote{We also note the successful observation of fractional Chern insulating phases in graphene based heterostructures under strong physical magnetic fields \cite{Spanton:2017vf}.} Early achievements include the square lattice with staggered magnetic flux \cite{Aidelsburger:2011hl, 2010PhRvA..82f3625M} that was generated by suitably tailored laser-induced hoppings \cite{Jaksch:2003ud, Gerbier:2010ho}. More recently, experiments have achieved homogeneous magnetic flux using related approaches \cite{Aidelsburger:2013ew, 2013PhRvL.111r5302M, Chin:2013kc, Cooper:2018tr}. The Chern bands of the Haldane model \cite{Haldane:1988gh} were also successfully engineered using a lattice shaking approach \cite{Jotzu:2014kz}. Features of the non-trivial band single band topology have been successfully identified \cite{Duca:2015cz, Aidelsburger:2015hm}.
Another groundbreaking line of research has exploited spatially dependent dressed states of atoms in order to create a Berry phase emulating the Aharonov-Bohm effect of charged particles moving in a magnetic field \cite{Lin:2009us}. The experimental realization of this approach \cite{Lin:2009us} has prompted further theoretical developments in order to maximize the achievable flux density in so-called optical flux lattices \cite{Cooper:2011iv, Cooper:2011dq}. These systems rely on modulating the optical dressed states of multi-state atoms on the scale of the optical wavelength, thus accessing the smallest possible length scales for light-matter coupled systems, and provide a viable route to observe fractional quantum Hall physics \cite{Cooper:2013jg, Sterdyniak:2015jo, Sterdyniak:2015kp}. Experimental progress has been reported on the intimately related case of emulating spin-orbit coupling in two dimensional systems \cite{Wu:2016kv, Huang:2016kf, Sun:2017wa}.

So far, attempts to emulate optical lattices with synthetic gauge fields have focused on continuum gases or on simple optical lattice geometries such as square and triangular lattices \cite{Windpassinger:2013is}. However, optical lattices without gauge fields have already been demonstrated for more complex geometries such as the kagome lattice \cite{Jo:2012br}, which is achieved by removing \emph{sites} from an underlying triangular lattice. 
Lattice geometry plays a particularly important role in the presence of magnetic flux, as it can affect the single particle spectrum dramatically. Indeed, the elegant Hofstadter butterfly seen in the spectrum of the square lattice \cite{Hofstadter:1976wt} is strongly altered in other geometries such as the triangular \cite{Claro:1979tn} or hexagonal lattices \cite{Rammal:1985ef}. This provides a strong incentive to achieve synthetic gauge fields in a number of different lattice geometries.

It is well understood how complex lattice geometries can be realised in scalar optical lattices by exploiting the superposition of several optical lattice potentials \cite{Jo:2012br,Tarruell:2012db,Uehlinger:2013dh}. In this paper, we explore how this design principle can be extended to create optical flux lattices with non-trivial connectivity by superposing scalar sub-/superlattice potentials to an optical flux lattice that generates non-trivial Berry phases from adiabatic motion within the space of internal states of the trapped atoms. We demonstrate that a scalar potential may be used to either remove \emph{bonds} or \emph{sites} from an underlying optical flux lattice of simpler geometry, as well as to split individual sites into multiple wells, all the while keeping the synthetic field intact. The basic principle for controlling bonds can be understood from a tight-binding picture: the dynamics of atoms in an optical lattice arises from hopping processes between local Wannier states that are localized in the minima or wells of the optical potential \cite{Jaksch:1998ee}. The amplitude of hopping processes is given by the overlap of these wave functions. As the overlap is dominated by the exponential tails penetrating the potential maxima that separate adjacent wells, hopping is extremely sensitive to the magnitude of this potential. Therefore, hopping can be almost completely suppressed by increasing the height of the potential maximum between two wells when a scalar potential is added at those locations. Generally, we wish to suppress bonds on a periodic sublattice of an underlying optical flux lattice, so this can be implemented by superposing an additional scalar optical lattice potential which acts equally on all internal states. In practice, optical flux lattices operate in an intermediate coupling regime where the lattice potential is sufficiently shallow for atoms to occupy any position in space. One of the main results of the current work is to demonstrate that complex optical flux lattices can operate in a regime of weak coupling that remains far from the tight-binding limit: we provide a specific example showing that the dispersion of the tight-binding picture is reproduced closely even in the regime of shallow lattice depth with potential depth of order of the atomic recoil energy.

In order to demonstrate our general principle, we propose and analyse in detail a new realization for synthetic fields in the dice lattice (also known as $\mathcal{T}_3$-lattice) where the specific flux density of $\Phi=\Phi_0/2$ per plaquette yields a particularly surprising band structure with three pairs of perfectly flat bands that conserve time-reversal symmetry \cite{Vidal:1998gx}. The flat bands and compactly localized single particle states found in this lattice are caused by a phenomenon of destructive interference known as Aharonov-Bohm caging \cite{Vidal:1998gx}. This regime would be particularly well suited to reach interesting correlation phenomena \cite{Korshunov:2005iq,2012PhRvL.108d5306M,Payrits:2014ft}, but previous proposals for synthetic fields in a dice lattice geometry that have focused on a different regime with dispersive Chern bands \cite{Andrijauskas:2015em}.
Unlike most flat band models achievable in cold atoms \cite{Huber:2010bc}, the flat bands of the $\pi$-flux dice lattice model are fully gapped. Owing to the flatness of the band dispersion, even weak interactions give rise to exotic phases in the dice lattice model, including a superfluid phase in the half filled lowest band \cite{2012PhRvL.108d5306M} as well as highly degenerate vortex lattice configurations at larger density \cite{Korshunov:2005iq, 2012PhRvL.108d5306M} that provide a playing field for order-by-disorder phenomena \cite{Payrits:2014ft}. Hence, akin to the physics of flat band ferromagnetism \cite{Mielke:1999jk, Noda:2014ku}, the dominant phases in the dice lattice provide interesting alternatives to more conventional features of Bose condensation in dispersive bands \cite{2010PhRvA..82f3625M}.

To further contrast the new proposal with more conventional techniques, we also present an alternative design for a dice lattice with a synthetic $\pi$-flux based on alkaline earth atoms trapped by light near their anti-magic wavelength. We describe a set-up creating laser-induced hoppings according to the connectivity of the dice lattice, that can be realised using far-detuned transitions following Ref.~\cite{Gerbier:2010ho}. Our design explicitly constructs the tight-binding Hamiltonian within the magnetic unit cell, containing a total of six sites, which is repeated due to the inherent periodicity of the trapping lasers. We find that the two designs involve similar number of laser sources, and we argue that requirements on phase stability favour the optical flux lattice approach.

The paper is organised as follows. In section \ref{sec:background}, we review how the concept of adiabatic motion in optical dressed states enables the creation of optical flux lattices, and we establish our notations. In section \ref{sec:EliminatingBonds}, we introduce the idea of changing lattice connectivity by removing bonds from an optical flux lattice at the level of a tight binding approach, and perform an analysis of its translational symmetries. In section \ref{sec:DiceLattice} we detail how the idea can be exploited to realize the dice lattice geometry with half a flux quantum per plaquette, focusing on a tight-binding picture.
Section \ref{sec:BeyondTightBinding} gives the general formalism for studying optical flux lattices beyond the tight-binding limit in reciprocal space, and we use the example of the dice lattice geometry to demonstrate the role of spin-translation symmetries of the flux lattice Hamiltonian.
In section \ref{sec:DiceQuantitative}, we provide detailed calculations of the band structure for realistic parameters in our dice flux lattice geometry, focusing on the limit of a shallow lattice. 
Section \ref{sec:DiceTightBinding} provides the alternative design, based on laser-induced hoppings in a deep optical lattice, and we conclude in section \ref{sec:Conclusions}.

\section{Background: Optical Flux Lattices}
\label{sec:background}

The optical flux lattice approach is motivated by the principle of adiabatic motion of atoms, such that they remain in their local ground-state $| \Psi(\br)\rangle $ along their trajectory $\br(t)$ \cite{Cooper:2011iv}. Upon completion of a closed path $\mathcal{C}$, the wavefunction of the atoms acquires a geometrical Berry phase $\gamma = \oint_\mathcal{C} q\mathbf{A} d\mathbf{l}$, given by the line integral over the (real space) Berry connection $q\mathbf{A} = i\hbar \langle \Psi | \bm{\nabla} \Psi \rangle$ (with a fictitious charge $q$) \cite{Lin:2009us}. This geometric phase mimics the Aharonov-Bohm coupling of a charged particle to the vector potential of a physical magnetic field, which has the same form. It also useful to think of the corresponding flux density $n_\phi=q/h(\bm{\nabla} \times \mathbf{A})\cdot\unit{e}_3$.

The presence of vortices in the Berry connection allows one to achieve flux densities of order one magnetic flux quantum per unit cell of the optical flux lattice. Here, we will consider as our starting point the explicit example of the triangular flux lattice of Ref.~\cite{Cooper:2011iv} for a two-state system with the Hamiltonian
\be
\label{eq:Htriangle}
\hat \mathcal{H}_\bigtriangleup = \frac{\mathbf{\hat p}^2}{2m} \mathbf{1} + \mathcal{V} \mathbf{M}(\br)\cdot \bm{\hat \sigma},
\ee
where $\mathbf{1}$ is the $2\times 2$ identity matrix in spin-space, $\bm{\hat \sigma} = (\hat \sigma_1, \hat \sigma_2, \hat \sigma_3)$ is the vector of Pauli matrices, and $\mathcal{V}$ is the depth of the optical lattice. We consider the triangular optical lattice potential described by
\be
\label{eq:BlochVector}
\mathbf{M}(\br) = \cos (\cvec{\kappa}_1 \br) \unit{e}_1 + \cos (\cvec{\kappa}_2 \br) \unit{e}_2 + \cos (\cvec{\kappa}_3 \br) \unit{e}_3,
\ee
where $\unit{e}_i$ are the cartesian unit vectors, and the wave vectors $\cvec{\kappa}_1 = (1,0)\kappa$, $\cvec{\kappa}_2=(1/2,\sqrt{3}/2)\kappa$, and $\cvec{\kappa}_3=(-1/2,\sqrt{3}/2)\kappa$  are chosen to yield a lattice potential with minima separated by a lattice vector $a$, i.e., we require $\kappa=\frac{2\pi}{\sqrt{3}a}$. In our notations, we highlight constant vectors defined by externally imposed geometrical features such as $\cvec{\kappa}_i$ in bold-face with an additional arrow, while vectors representing variables like $\br$ are denoted in simple bold font. Note that specific implementations of a triangular optical flux lattice such as (\ref{eq:Htriangle}) may be realised by various optical coupling schemes. Detailed implementations have been presented elsewhere (see, e.g., Ref.~\cite{Cooper:2011dq}), so we shall work with the simplest model in the current paper.

\begin{figure}[tt]
\begin{center}
\includegraphics[width=0.95\columnwidth]{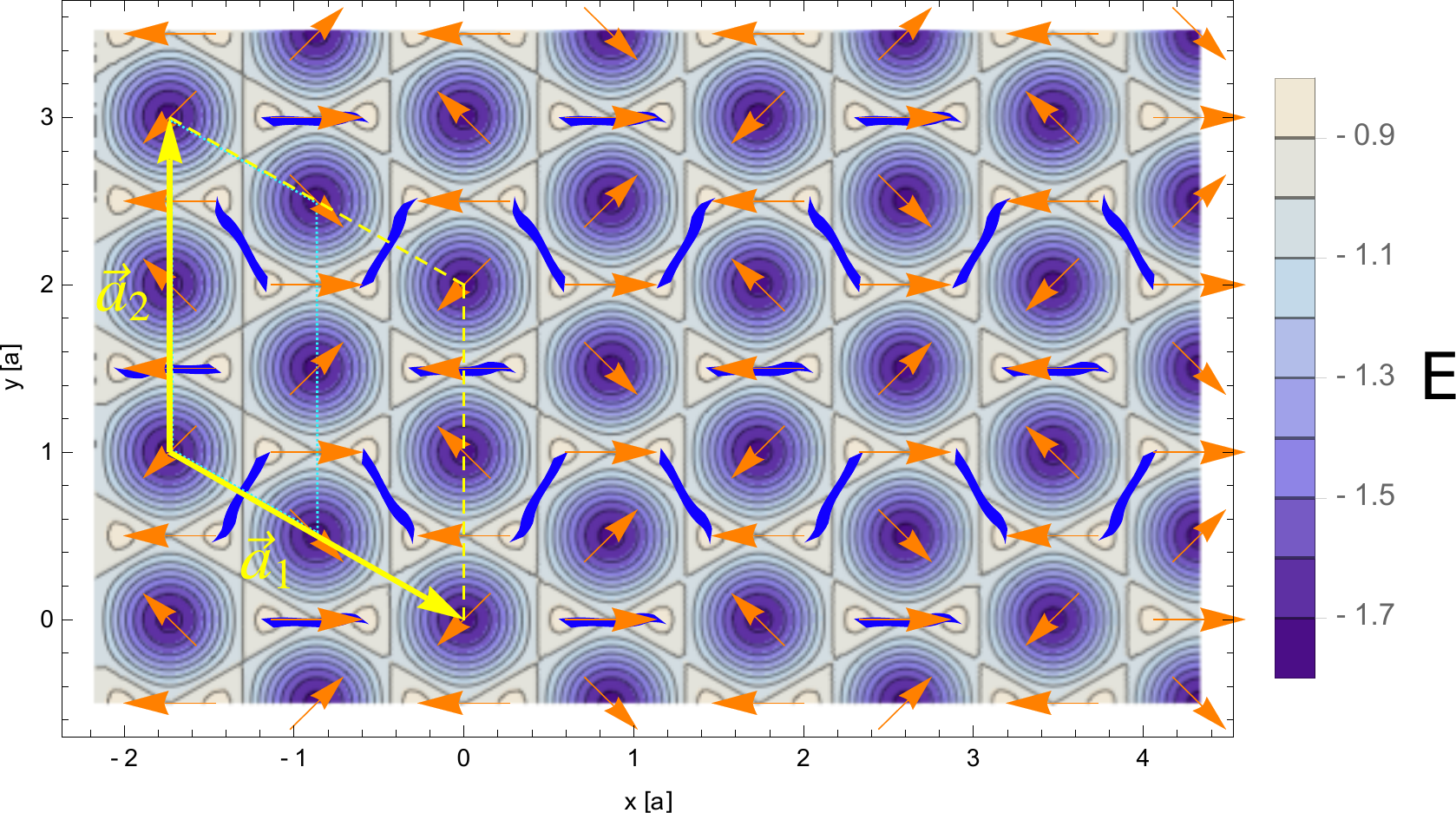}
\caption{Contour plot of the energy landscape for the triangular optical flux lattice with two flux quanta per unit cell of Ref.~\cite{Cooper:2011iv}, the starting point for our construction. Orange arrows show the in-plane components of the local Bloch vector. The unit cell is spanned by the vectors $\cvec{a}_1$, $\cvec{a}_2$, contains $4$ triangular lattice sites, and encloses $2$ flux quanta. Thanks to a spin-translation symmetry, this can be reduced to a reduced unit cell of size $[\cvec{a}_1/2$, $\cvec{a}_2]$ (dotted cyan lines). In this paper, we show how this flux lattice can be modified to yield an optical dice flux lattice by eliminating bonds: a dice lattice is obtained by impeding tunnelling across the links which are crossed out with blue wavy lines.}
\label{fig:TriangularFL}
\end{center}
\end{figure}

In the adiabatic limit $m\to \infty$, it is easily checked that the Hamiltonian (\ref{eq:Htriangle}) has eigenvalues $E_\pm(\br) = \pm \mathcal{V} | \mathbf{M} |$, and the local Bloch vector for the lower band, $\mathbf{\hat  n} = \langle \Psi_-(\br)|\bm{\hat \sigma} | \Psi_- (\br) \rangle $, is simply given by the direction of $-\mathbf{M}$, i.e., $\mathbf{\hat n} = -\mathbf{\hat M} \equiv -\mathbf{M} / |\mathbf{M}|$. The states $|\Psi_\pm\rangle$ are also the eigenstates for the class of Hamiltonians $\hat \mathcal{H}' = \hat \mathcal{H}_\bigtriangleup + \hat \mathbf{V}_\mathrm{s}$, for arbitrary scalar (i.e., spin-independent) potentials $\hat \mathbf{V}_\mathrm{s}(\br) = V_\mathrm{s}(\br) \hat \mathbf{1}$. The energy landscape for the unperturbed triangular flux lattice (\ref{eq:Htriangle}) is shown in Fig.~\ref{fig:TriangularFL}. Note that the unit cell of this lattice encloses two flux quanta within an area containing four local minima of the energy, which we can think of as four lattice sites in the tight-binding limit of a deep optical flux lattice \cite{Cooper:2011iv}. For our choice of units, the lattice vectors spanning the unit cell are given by $\cvec{a}_1 = (\sqrt{3},-1)a$, and $\cvec{a}_2 = (0,2)a$, as highlighted in Fig.~\ref{fig:TriangularFL}.

The periodicity of the energy landscape suggests that the Hamiltonian (\ref{eq:Htriangle}) has a higher translational symmetry than that by the above-mentioned lattice vectors $\cvec{a}_i$. While energetically equivalent, the eigenstates at the four energy minima in the unit cell are distinct. However, the higher symmetry of the Hamiltonian can be revealed by generalized translation operators that incorporate a rotation in spin space \cite{Cooper:2011iv}. Available spin-translation operators are
\begin{eqnarray}
\label{eq:T12}
\hat T_1 = \hat \sigma_2 e^{\frac{1}{2}\cvec{a}_1\cdot \bm{\nabla}},\qquad\; \hat T_2 = \hat \sigma_1 e^{\frac{1}{2}\cvec{a}_2\cdot \bm{\nabla}},
\end{eqnarray}
with $[\hat T_i, \hat \mathcal{H}_\bigtriangleup] = 0$ ($i=1,2$), but  $[\hat T_1, \hat T_2] \neq 0$. Nonetheless, we find that $[\hat T_1, \hat T_2^2] = 0 $, so we can classify the eigenvalues of $\hat \mathcal{H}_\bigtriangleup$ with the quantum numbers of both $\hat T_1$, and $\hat T_2^2 = \exp(\cvec{a}_2\cdot \bm{\nabla}) \equiv \hat K(\cvec{a}_2)$, as the latter reduces to a regular translation $\hat K(\cvec{a}_2)$ by $\cvec{a}_2$. For a detailed discussion of these symmetry operations in the triangular lattice, see Ref.~\cite{Cooper:2011iv}.

\section{Changing Lattice Topology via Scalar Potentials}
\label{sec:EliminatingBonds}
In the deep optical lattice limit, we can consider optical flux lattices as tight binding models where motion between two `sites' or local minima of the energy landscape is described by a tight binding model with complex hopping elements.  We now examine how a change in the lattice topology emulated by optical flux lattices is achieved either by `removing sites' or by `removing bonds' in this tight binding model, as was already achieved for scalar optical lattices \cite{Tarruell:2012db,Jo:2012br}. As we will demonstrate below, this idea can indeed also be realised in optical flux lattices by applying an additional scalar optical lattice potential to either suppress lattice sites or the connectivity between them, while the distribution of flux generated by the underlying optical flux lattice is kept intact. 

\begin{figure}[t]
\begin{center}
\includegraphics[width=0.95\columnwidth]{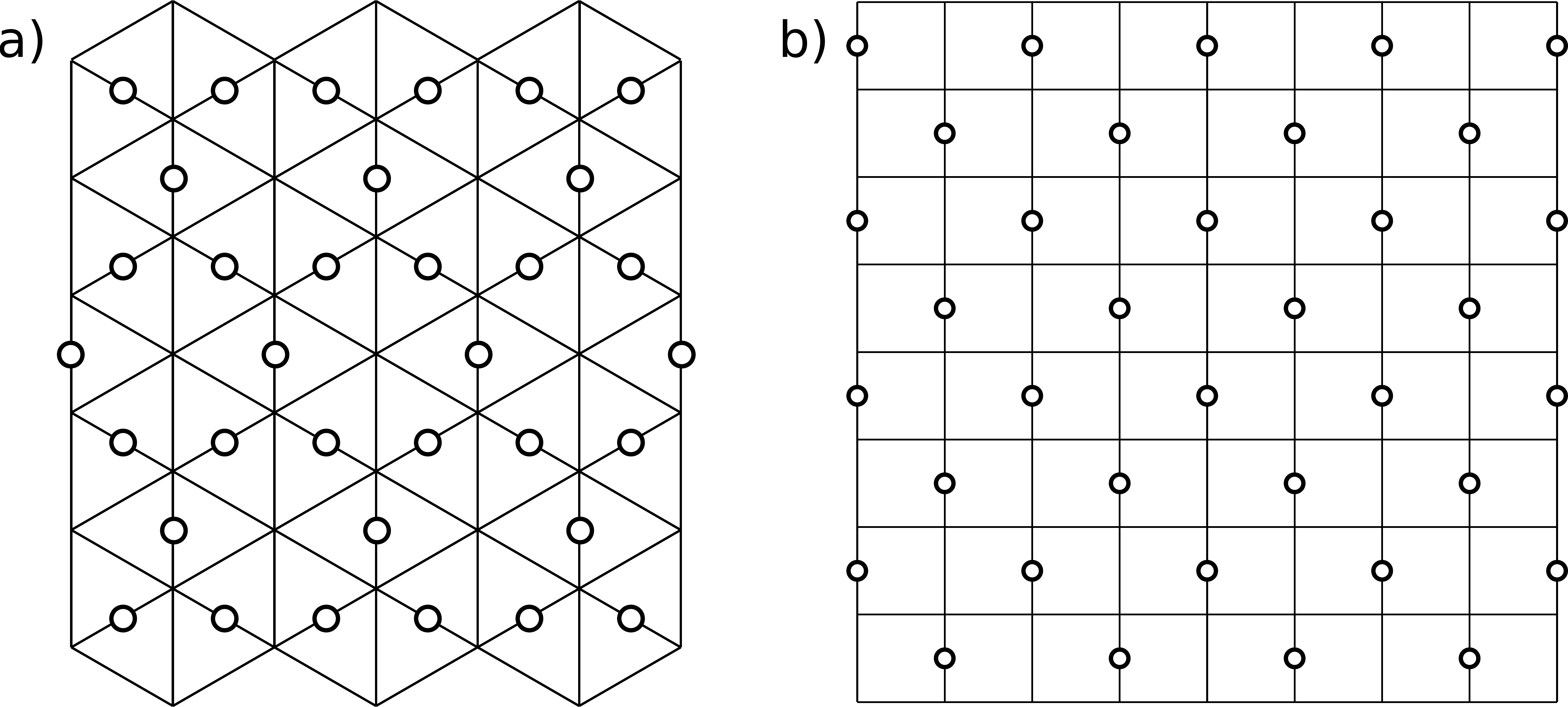}
\caption{Examples of new lattice topologies that emerge by elimination of bonds from an underlying graph, where suppressed hoppings are symbolized as open circles. The triangular lattice (a) can be reduced to a dice lattice. Here, the centers of eliminated bonds form a kagome lattice. A square lattice (b) can be reduced to a brickwork lattice which has the connectivity of a honeycomb lattice. Here, the centers of eliminated bonds again form a square lattice. A regular honeycomb lattice can also be recovered from this set-up by scaling the $x$-axis by one half.}
\label{fig:ExampleGraphs}
\end{center}
\end{figure}

Some examples of cutting bonds are visualized in Fig.~\ref{fig:ExampleGraphs}. There are already similar experimental realisations of tuneable optical lattices obtained by superposing multiple standing waves \cite{Tarruell:2012db,Jo:2012br}. An additional consideration for flux lattices arises in the tight-binding limit, where flux through each plaquette is defined only modulo $2\pi$. As the elimination of links joins the two adjoining plaquettes into a single one, this construction yields non-trivial flux lattices only if the total flux in the resulting merged plaquette is not an integer multiple of the flux quantum $\Phi_0$. Similarly, the removal of sites merges several adjoining plaquettes, so the same consideration applies. For example, a hexagonal lattice can be obtained by removing a sublattice of sites of an underlying triangular lattice. In this case, six neighbouring triangular plaquettes are joined into a hexagonal one, so this yields non-trivial results if the flux per triangular plaquette is not a multiple of $\Phi_0/6$.

\section{Case Study: the Dice Lattice}
\label{sec:DiceLattice}
For the remainder of this paper, we focus on a case study of eliminating bonds in a triangular flux lattice. Alongside the elementary unit cell of the flux lattice, Fig.~\ref{fig:TriangularFL} highlights the bonds that need to be severed in the triangular lattice so as to reduce its connectivity to a dice lattice geometry%(also known as the $\mathcal{T}_3$ lattice)
. As shown more clearly in Fig.~\ref{fig:ExampleGraphs}(a), we find that mid-points of these bonds form a kagome lattice with lattice constant $a' = \sqrt{3}/2 a$. From Fig.~\ref{fig:TriangularFL}, it is also clear that the pattern of eliminated bonds has a different periodicity as the unit cell $[\cvec{a}_1, \cvec{a}_2]$ of the triangular optical flux lattice. This will be further discussed, below.

In our cold atom realization of an optical dice flux lattice, the maxima of an additional scalar optical potential are aligned with the centre points of the bonds of an underlying triangular optical flux lattice. 
As experiments by the Stamper-Kurn group demonstrate, an attractive kagome optical lattice can be achieved by combining a blue-detuned (i.e., regions of high intensity are repulsive) short-wavelength triangular optical lattice with a red-detuned (attractive) triangular lattice of twofold lattice constant \cite{Jo:2012br}. Experimentally, it is difficult to keep these two lattices in register, but this challenges has been successfully addressed \cite{Jo:2012br}.
Here, we require a \emph{repulsive} kagome lattice, which is rotated by $\pi/6$ with respect to an underlying triangular optical flux lattice (\ref{eq:Htriangle}), again implying that the two light potentials have to be kept in phase as in the kagome lattice realisation of \cite{Jo:2012br}. The corresponding optical potential is formed by a red-detuned short-wavelength scalar optical lattice $V_\mathrm{SW}$ at wave number $\kappa^\perp=2\kappa/\sqrt{3}$, as well as a blue-detuned long-wavelength scalar superlattice $V_\mathrm{LW}$ with wave number $\kappa^\perp/2$. The full Hamiltonian of our optical dice flux lattice is then obtained by superposing all three components
\be
\label{eq:DiceFL}
\hat \mathcal{H}_\mathrm{dice}(r, b) = \hat \mathcal{H}_\bigtriangleup + \left[r V_\mathrm{SW}(\hat \br) + b V_\mathrm{LW}(\hat \br)\right] \mathbf{1}.
\ee
Here, the parameters $b>0$ and $r<0$ give the amplitude of the scalar beams relative to the spin-dependent fields, and the explicit form of the required short- and long-wavelength potentials are given by
\begin{eqnarray}
\label{eq:VsL}
V_\mathrm{SW} (\br) &= \mathcal{V}
\left[ \sin^2\left(\cvec{\kappa}_1^\perp \br\right)  + \sin^2\left(\cvec{\kappa}_2^\perp \br\right) + \sin^2\left(\cvec{\kappa}_3^\perp \br\right)\right]
\end{eqnarray}
for the red detuned beam that is attractive, and that should thus contribute with an amplitude $r<0$, and
\begin{eqnarray}
\label{eq:VsSL}
V_\mathrm{LW} (\br) &= \mathcal{V} \left[ \sin^2\left(\frac{\cvec{\kappa}_1^\perp}{2} \br\right)  + \sin^2\left(\frac{\cvec{\kappa}_2^\perp}{2} \br\right) + \sin^2\left(\frac{\cvec{\kappa}_3^\perp}{2} \br\right)\right]
\end{eqnarray}
for the blue detuned beam that should provide a repulsive potential with an amplitude $b>0$, and $\cvec{\kappa}_i^\perp = 2/\sqrt{3} \hat e_3 \wedge \cvec{\kappa}_i$ throughout. Note that both these contributions are scalar, i.e., they are diagonal in spin space. 
In the adiabatic limit (i.e., disregarding kinetic energy), the local energy eigenvalues are readily obtained as $E_\pm^\mathrm{dice}(r,b) = \pm \mathcal{V} |\mathbf{M}| +  r V_\mathrm{SW}(\br) +  b V_\mathrm{LW}(\br)$, and the local eigenstates are unchanged with respect to the triangular optical flux lattice.

\begin{figure*}[t]
\begin{center}
\includegraphics[width=0.95\columnwidth]{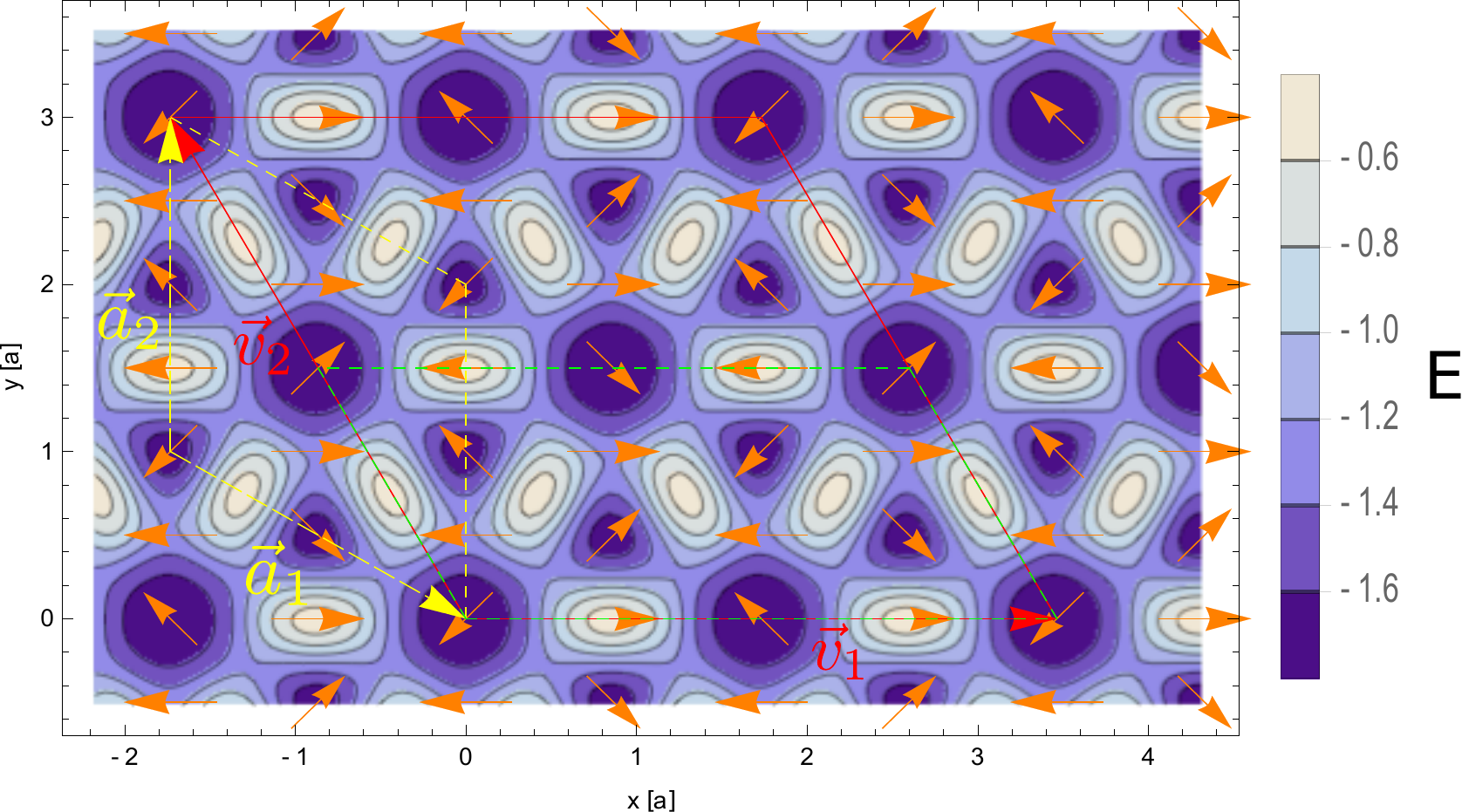}
\caption{Contours show the adiabatic energy landscape of the dice flux lattice (\ref{eq:DiceFL}) with $-r=b=1/8$, obtained by knocking out bonds from the underlying triangular optical flux lattice shown in Fig.~\ref{fig:TriangularFL}. The periodicities of the lattice result from a combination of the periodicity in the energy landscape (shown as a density plot with minima in dark blue) and the local Bloch vectors ($x$-$y$-components shown as orange arrows). The original unit cell $[\cvec{a}_1,\cvec{a}_2]$ is highlighted in dashed yellow lines/arrows. As the scalar potential has different periodicity than the flux lattice, the elementary unit cell of the dice flux lattice is enlarged and contains $12$ sites. The figure shows the dice unit cell in red full lines, spanned by vectors marked as $[\cvec{v}_1, \cvec{v}_2]$. Thanks to a combined symmetry of spin rotation and translations (see main text), the unit cell can be reduced to half that size, shown as the region $[\cvec{v}_1, \cvec{v}_2/2]$ enclosed in dashed green lines.}
\label{fig:DiceFluxLattice}
\end{center}
\end{figure*}

Let us now discuss the symmetries of the optical dice flux lattice Hamiltonian. As we noted previously, it does not have the full translational symmetry of the triangular optical flux lattice. The resulting situation is best discussed in terms of Fig.~\ref{fig:DiceFluxLattice}, which shows the energy landscape (contours; darker blue indicates minima), as well as the $x$-$y$-components of the local Bloch vector (orange arrows).
In the presence of the scalar potentials (\ref{eq:VsL}, \ref{eq:VsSL}), the energy landscape contains lattice sites with three different profiles: the most prominent minima form the `hubs' or sixfold connected sites of the dice lattice, such as the one at the origin $\br=(0,0)$. They are surrounded by six smaller minima, the `rims' or threefold connected sites. These are slightly triangular and can be either pointing upwards [such as at $\br=(\sqrt{3}/2,1/2)a$] or downwards [as at $\br=(0,1)a$]. In addition, lattice sites differ in terms of the spin-content of the local wavefunction. Looking at the in-plane components of the local Bloch-vectors, it is apparent that a fundamental unit cell of our optical dice flux lattice is enclosed by the vectors marked in Fig.~\ref{fig:DiceFluxLattice} as $\cvec{v}_1 = (2\sqrt{3},0)a$, and $\cvec{v}_2 = (-\sqrt{3},3)a$, which connect hubs with identical Bloch vectors. Due to the distinct periodicities, this unit cells contains 12 sites of the underlying triangular lattice so it is enlarged threefold with respect to the unit cell of the original triangular optical flux lattice.

The Hamiltonian (\ref{eq:DiceFL}) contains an additional symmetry, which can be constructed in terms of the spin-translation operators $\hat T_{1,2}$  in (\ref{eq:T12}). Let us construct suitable spin-translations $\hat S_{1,2}$ along the half lattice vectors $\frac{1}{2}\cvec{v}_{1,2}$. These can be expressed in terms of $\hat T_{1,2}$ as:
\begin{eqnarray}
\hat S_1 &= \hat T_1^2 \hat T_2^{\phantom{-1}} = \hat \sigma_1 \, e^{\frac{1}{2}(2\cvec{a}_1+\cvec{a}_2) \cdot \bm{\nabla}},\\
\hat S_2 &= \hat T_1^{-1} \hat T_2 = i \hat \sigma_3 \, e^{\frac{1}{2}(-\cvec{a}_1+\cvec{a}_2)\cdot \bm{\nabla}}.
\end{eqnarray}

We note that both $\hat S_i$ commute with the Hamiltonian, i.e.~$[\hat S_{1,2},\hat \mathcal{H}_\mathrm{dice}]=0$. Furthermore, their squares are simple translations, which confirms that we have chosen the unit cell correctly. For instance, $\hat S_1^2 = \hat \sigma_1^2 e^{(2\cvec{a}_1+ \cvec{a}_2)\cdot \bm{\nabla}} = e^{\cvec{v}_1\cdot \bm{\nabla}}$, which equals a pure translation $\hat K(\cvec{v}_1)$ under the lattice vector $\cvec{v}_1$. However, the translations $\hat S_1$ and $\hat S_2$ do not commute with each other, as $[\hat S_1,\hat S_2]\neq 0$. Given that $\hat S_2$ is diagonal in spin-space, we select this operator as our supplementary symmetry in formulating the single-particle Hilbert-space, and we can then use the eigenvalues of the set of commuting operators $\hat \mathcal{H}$, $\hat S_1^2$, and $\hat S_2$ to label eigenstates. This results in a reduced unit cell in real space, spanned by $[\cvec{v}_1, \cvec{v}_2/2]$, as shown in green dotted lines in Fig.~\ref{fig:DiceFluxLattice}, such that eigenstates in the remainder of the full unit cell can be recovered by applying $\hat S_2$ to their symmetry related points in the reduced cell.

\section{Spin-Translation Symmetry in Shallow Flux Lattices}
\label{sec:BeyondTightBinding}

The arguments of the preceding section can be placed on a more robust foundation by considering the full Hamiltonian of the flux lattice beyond the tight binding limit, i.e., including kinetic energy. In order to capture the effect of the kinetic energy term, it is convenient to study the flux lattice Hamiltonian as a tight-binding model in reciprocal space \cite{Cooper:2012bt}. Here, we review and extend this formalism to take into account the spin-translation symmetries, as realised by the operators $\hat S_1$, $\hat S_2$ identified above.

\subsection{General Formalism}

Having identified the periodicity of the problem in the real-space unit-cell (UC) spanned by $[\cvec{v}_1,\cvec{v}_2]$, we know that the wave functions in reciprocal space are defined on a fundamental Brillouin zone (BZ) spanned by the reciprocal lattice vectors
\be
\label{eq:reciprocalLV}
\cvec{g}_i = \epsilon_{ij}\frac{2\pi  \,  \bm{\cvec{v}_j \wedge \hat e}_3}{(\cvec{v}_1 \wedge \cvec{v}_2)\cdot \bm{\hat e}_3},\quad i=1,2
\ee
where $\epsilon_{ij}$ is the totally anti-symmetric tensor. The reciprocal lattice vectors thus satisfy $\cvec{g}_i \cdot \cvec{v}_j=\delta_{i,j}$. Now, let us turn to discuss the momentum transfers, which are obtained as the matrix elements of the interaction $\mathbf{V}(\br)$ in the basis of plane-wave states with $\langle \br |\bk,\alpha\rangle = e^{i\bk\cdot \br}\otimes|\alpha\rangle$ for the spin component $\alpha$. One finds that the matrix elements depend only on the momentum transfer $\Delta\bk = \bk'-\bk$
\begin{eqnarray}
V_{\bk',\bk}^{\alpha'\alpha} = V_{(\bk'-\bk)}^{\alpha'\alpha} = \langle \bk',\alpha' | \mathbf{V}(\br) |\bk,\alpha\rangle.
\end{eqnarray}

\begin{figure*}[t]
\begin{center}
\includegraphics[width=0.92\textwidth]{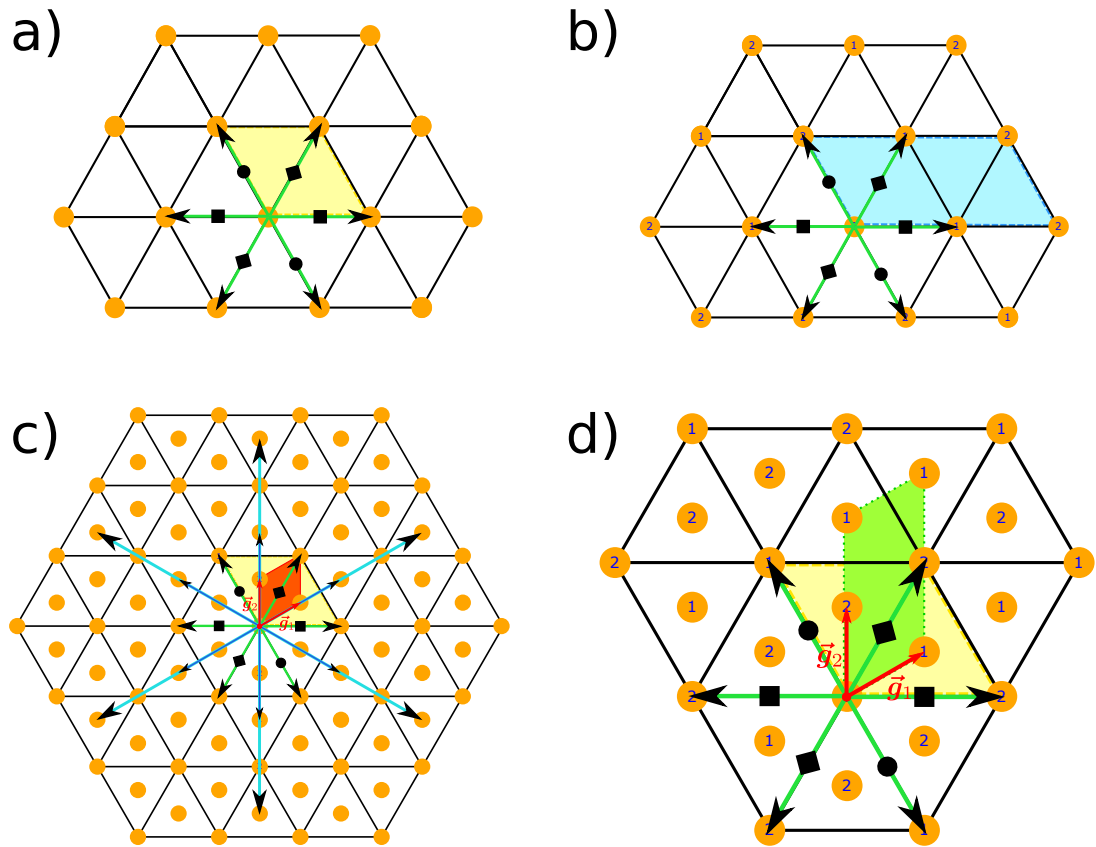}
\caption{Representation of optical flux lattices as tight binding models on a grid of $k$-points (circles) highlighting momentum transfers, or ``hoppings'', induced by absorption/emission of photons (arrows). We show the lattice of accessible momentum transfers for the triangular (a,b) and dice-lattice (c,d) geometries. Note the panels are scaled differently, with the links shown in black corresponding to the same momentum transfer throughout.
(a) The lasers of the triangular optical flux lattice (\ref{eq:Htriangle}) propagate along directions $\cvec{\kappa}_1\propto (1,0)^t$, $\cvec{\kappa}_2\propto (1/2, \sqrt{3}/2)^t$, and $\cvec{\kappa}_3\propto (1/2, \sqrt{3}/2)^t$. These momentum transfers induce spin-transitions given by $\hat \sigma_1$, $\hat \sigma_2$ and $\hat \sigma_3$ respectively, highlighted by squares, diamonds, and circles on the corresponding arrows. The fundamental Brillouin zone is shaded in yellow. (b) Taking into account the spin-translation symmetry of (\ref{eq:Htriangle},\ref{eq:BlochVector}), one can assign a definite spin-state to the accessible $k$-points (denoted as 1 or 2 in the figure), while the corresponding enlarged Brillouin zone (blue shade) is doubled along $\hat \mathbf{e}_1$.
(c) The reciprocal-space representation of the optical dice flux lattice includes the triangular lattice transitions, as well as additional momentum transfers due to the scalar potential $V_\mathrm{kag} = rV_\mathrm{SW}+b V_\mathrm{LW}$ of Eq.~(\ref{eq:VsL},\ref{eq:VsSL}). These ``hoppings'' along directions $\cvec{\kappa}_i^\perp$ connect to additional $k$-points located in the centers of the original triangular lattice, yielding a Brillouin zone for the dice lattice (red shade) that is $1/3$ the size of the Brillouin zone for the triangular lattice (yellow shade). The reciprocal lattice vectors $\cvec{g}_1$, $\cvec{g}_2$ are shown as red arrows. (d) The  spin-translation symmetry of the optical dice flux lattice again leads to a unique labelling of spin states  $1,2$ for all possible momentum transfers. This yields an enlarged Brillouin zone, which is stretched along $\cvec{g}_2$ and covers the region $[\cvec{g}_1,2\cvec{g}_2]$ (green shade).
}
\label{fig:ReciprocalDiceFluxLattice}
\end{center}
\end{figure*}

According to Bloch's theorem, eigenstates $|\Psi_{n\mathbf{q}}\rangle $ are uniquely labelled by a band index $n$ and momentum $\mathbf{q}$ in the first Brillouin zone, while larger momenta can be decomposed as $\bk = \mathbf{q} + \mathbf{G}$  into a part lying in the BZ and a reciprocal lattice vector $\mathbf{G}_{st}=s \cvec{g}_1 + t \cvec{g}_2$ with $s$, $t$ integer. In its Bloch form the wavefunction reads
\be
\label{eq:BlochForm}
|\Psi_{n\mathbf{q}}\rangle  = \sum_\alpha u_{n\mathbf{q}}^\alpha (\br) | \mathbf{q}, \alpha \rangle \equiv \sum_{\alpha,s,t} c_{n\mathbf{q},\mathbf{G}_{st}}^\alpha  |\mathbf{q}+\mathbf{G}_{st}, \alpha \rangle,
\ee
with expansion coefficients $c_{n\mathbf{q},\mathbf{G}}^\alpha$. As was noted previously \cite{Cooper:2012bt}, the flux lattice Hamiltonian takes the form of a tight binding model in reciprocal space in which the kinetic energy plays the role of a harmonic confinement:
\be
\label{eq:ReciprocalTightBinding}
\hat \mathcal{H}_\mathbf{q} = \sum_{\alpha,\mathbf{G}}\frac{\hbar^2 (\mathbf{q}+\mathbf{G})^2}{2m} \hat a^\dagger_{\alpha,\mathbf{q}+\mathbf{G}} \hat a_{\alpha,\mathbf{q}+\mathbf{G}} + \sum_{\alpha \alpha',\mathbf{G}\mathbf{G}'} V^{\alpha' \alpha}_{\mathbf{G'}-\mathbf{G}} \hat a^\dagger_{\alpha',\mathbf{q}+\mathbf{G}'} \hat a_{\alpha,\mathbf{q}+\mathbf{G}},
\ee
written here in terms of the annihilation (creation) operators $\hat a^{(\dagger)}_{\alpha,\bk}$ for the plane-wave basis.
We should also carefully note that all hoppings in this momentum-space tight-binding representation are relative to the wave-vector $\mathbf{q}$, hence they represent a lattice of achievable \emph{momentum transfers}, while in the usual case of tight-binding models in real space one is used to consider a lattice of fixed positions.

The depth of the optical lattice potential is reflected by the magnitude $\mathcal{V}$ of the largest entries in $V^{\alpha' \alpha}_{\Delta\bk}$. The typical kinetic energy is of order of the recoil energy, which we define as in terms of the relevant momentum transfer $\Delta \mathbf{p} = \hbar \Delta \mathbf{k}$ of the relevant laser beam as 
\be 
\label{eq:RecoilEnergy}
E_R= \frac{\hbar^2|\Delta \mathbf{k}|^2}{2m}.
\ee 
The adiabatic limit is recovered when $E_R \ll \mathcal{V}$, where the kinetic energy can be neglected and the problem is solved by Fourier transform back into real space, where position $\br$ plays the role of a conserved momentum. In the general case, (\ref{eq:ReciprocalTightBinding}) defines a matrix equation for the coefficients $c_{n\mathbf{q},\mathbf{G}}^\alpha$, which can be solved numerically as coefficients decay rapidly with the absolute value of momentum $|\mathbf{q}+\mathbf{G}|$.

\subsection{Role of Spin-Translation Symmetries in Complex Optical Flux Lattices}
The role of the spin-translation symmetries is more easily explained within an example.
Let us therefore focus on the reciprocal space picture of the dice flux lattice $\mathcal{H}_\mathrm{dice}$, that is illustrated in Fig.~\ref{fig:ReciprocalDiceFluxLattice}. % Version for reciprocal lattice vectors 1
For the components associated with the triangular flux lattice (\ref{eq:Htriangle}), we obtain the spin-dependent processes $\mathbf{\hat V}_{\cvec{\kappa}_1} = \mathcal{V}\hat \sigma_1$ with momentum transfer $\cvec{\kappa}_1 = -\cvec{g}_1+2\cvec{g}_2$, $\mathbf{\hat V}_{\cvec{\kappa}_2} = \mathcal{V}\hat \sigma_2$ with $\cvec{\kappa}_2 = \cvec{g}_1 + \cvec{g}_2$, and $\mathbf{\hat V}_{\cvec{\kappa}_3} = \mathcal{V}\hat \sigma_3$ with  $\cvec{\kappa}_3 = 2\cvec{g}_1 - \cvec{g}_2$, where the reciprocal lattice vectors $\cvec{g}_i$ are defined by the lattice vectors $\cvec{v}_i$ spanning the unit cell of the dice flux lattice according to (\ref{eq:DiceFL}). For later reference, note that these momentum transfers are proportional to the wave vectors of the three coupling lasers of the dice optical flux lattice, and are linear combinations in integer multiples of its reciprocal lattice vectors (\ref{eq:reciprocalLV}).

We display the momentum transfers of the underlying triangular optical flux lattice in Fig.~\ref{fig:ReciprocalDiceFluxLattice}(a), which also highlights the Brillouin zone corresponding to full the real-space unit cell $[\cvec{a}_1,\cvec{a}_2]$ of Fig.~\ref{fig:DiceFluxLattice}. Following \cite{Cooper:2012bt}, the spin-translation symmetry $\hat T_1$ of this model can be exposed by fixing the eigenvalue of the spin-translation operator, leading to a halving of the real space unit cell to $[\cvec{a}_1/2,\cvec{a}_2]$, thus doubling the Brillouin zone and leaving a definite spin state at each reciprocal lattice site, as shown in Fig.~\ref{fig:ReciprocalDiceFluxLattice}(b).

To obtain the dice optical flux lattice, we add to this picture the coupling to the scalar optical potentials generating the kagome lattice, Eqs.~(\ref{eq:VsL},\ref{eq:VsSL}), which contribute with momentum transfers corresponding to \emph{twice} their wave numbers, arising from the absorption of a photon from a standing wave laser followed by stimulated emission in the opposite direction. For $V_\mathrm{SW}$, we obtain momentum transfers $\Delta \cvec{k}_i^{SW}=2 \cvec{\kappa}_i^\perp$, with amplitude $\mathbf{\hat V}_{SW} = r\mathcal{V} \mathbf{\hat 1}$  and similarly for $V_\mathrm{LW}$ the momentum transfers are $\Delta \cvec{k}_i^{LW}=\cvec{\kappa}_i^\perp$ with amplitude $\mathbf{\hat V}_{LW} = b\mathcal{V} \mathbf{\hat 1}$. These momentum transfers are four- or twofold multiples of the reciprocal lattice vectors and their $\pi/3$ rotations. 

According to the enlarged unit cell in real space, the BZ of the dice lattice should cover one third of the area of the BZ for the triangular optical flux lattice. The corresponding lattice of possible momentum transfers is illustrated in Fig.~\ref{fig:ReciprocalDiceFluxLattice}(c), revealing a three times denser coverage of attainable $\mathbf{k}$-points.
The action of the spin-translation symmetry of the dice lattice model is again readily illustrated in this momentum space picture. Assume a single-particle wave-function has a non-zero amplitude for spin state 1 and vanishing amplitude for spin state 2 at momentum $\mathbf{q}$. Applying momentum- and spin-transfers to this initial state according to the tight-binding Hamiltonian (\ref{eq:ReciprocalTightBinding}), one can see that all related reciprocal lattice points at positions $\mathbf{q}+\mathbf{G}$ are reached with a definite spin quantum number. Equivalently, the Hamiltonian does not allow one to create any loops that return to the initial point with a different value of the spin. Choosing a spin state of $1$ at the central $\mathbf{k}$-point, one obtains the spin labels shown in Fig.~\ref{fig:ReciprocalDiceFluxLattice}(d). An equivalent labelling is obtained by interchanging labels `$1$' and `$2$' (or equivalently, by a translation of the figure under $\cvec{g}_2$). 

The spin-translation symmetry can be more formally derived from the eigenvalue equations of the spin-translation operators $\hat S_{1,2}$. We take $\hat S_2$ and $\hat S^2_1$ as the chosen symmetry generators commuting with the Hamiltonian, or $[\hat \mathcal{H},\hat S_1^2]= [ \hat \mathcal{H},\hat S_2] = [\hat S_1^2,\hat S_2]=0$, as discussed in Sec.~\ref{sec:DiceLattice}. This implies that the Hamiltonian is block-diagonal in the subspaces of fixed eigenvalues of $\hat S_1^2$, $\hat S_2$. Given the unitarity of these operators, we denote their eigenvalues as $\lambda_i=\exp(i \Theta_i)$, with $\hat S_1 ^2 |\Theta_1, \Theta_2\rangle = \exp (i \Theta_1) |\Theta_1, \Theta_2\rangle$ and $\hat S_2 |\Theta_1, \Theta_2\rangle = \exp (i \Theta_2) |\Theta_1, \Theta_2\rangle$, with $|\Theta_1, \Theta_2\rangle$ the corresponding eigenstates.
Consider then the explicit action of the generalised translations on momentum eigenstates
\begin{eqnarray}
\hat S_1^2 |\mathbf{k},\alpha\rangle = \mathbf{\hat 1}e^{i \cvec{v}_1 \cdot \mathbf{k}} |\mathbf{k},\alpha\rangle\nonumber, \\
\hat S_2 |\mathbf{k},\alpha\rangle = i\hat \sigma_3 e^{\frac{i}{2} \cvec{v}_2 \cdot \mathbf{k}} |\mathbf{k},\alpha\rangle.
\end{eqnarray}
We see that the phases are periodic under translations of $\mathbf{k} \to \mathbf{k}+\cvec{g}_1$ in the phase of $\hat S^2_1$, while the action of $\hat S_2$ is periodic under a doubled reciprocal lattice vector $\mathbf{k} \to \mathbf{k}+2\cvec{g}_2$, when the spin-state is fixed. Thus, we can label eigenstates by a momentum $\mathbf{q}$ taken to lie in the enlarged BZ $[\cvec{g}_1,2\cvec{g}_2]$ that is stretched twofold along the $\cvec{g}_1$-direction, as highlighted in Fig.~\ref{fig:ReciprocalDiceFluxLattice}(d). In this representation, each point of reciprocal momentum transfers can be assigned a definite spin state, as the momentum $\mathbf{q}$ in the enlarged BZ provides sufficient information to encode both the spin and momentum degrees of freedom. Alternatively, one could choose to represent the full range of possible eigenvalues $\Theta_2 \in [0,2\pi)$ by reducing the momentum to the fundamental Brillouin zone $[\cvec{g}_1,\cvec{g}_2]$, and recover the full range of $\Theta_2$ by taking into account both $\pm 1$ eigenvalues of the spin operator $\hat \sigma_3$.

\section{Quantitative Analysis of the $\pi$-flux Optical Dice Flux Lattice}
\label{sec:DiceQuantitative}

In this section, we provide a numerical study of the optical dice flux lattice introduced in section \ref{sec:BeyondTightBinding}. Numerics are performed in terms of the reduced unit cell $[\cvec{v}_1,\cvec{v}_2/2]$, or its reciprocal space counterpart. In other words, our implementation relies on resolving eigenstates of the generalized translations $\hat S_2$, as discussed above.

\begin{figure}[t]
\begin{center}
\includegraphics[width=0.9\columnwidth]{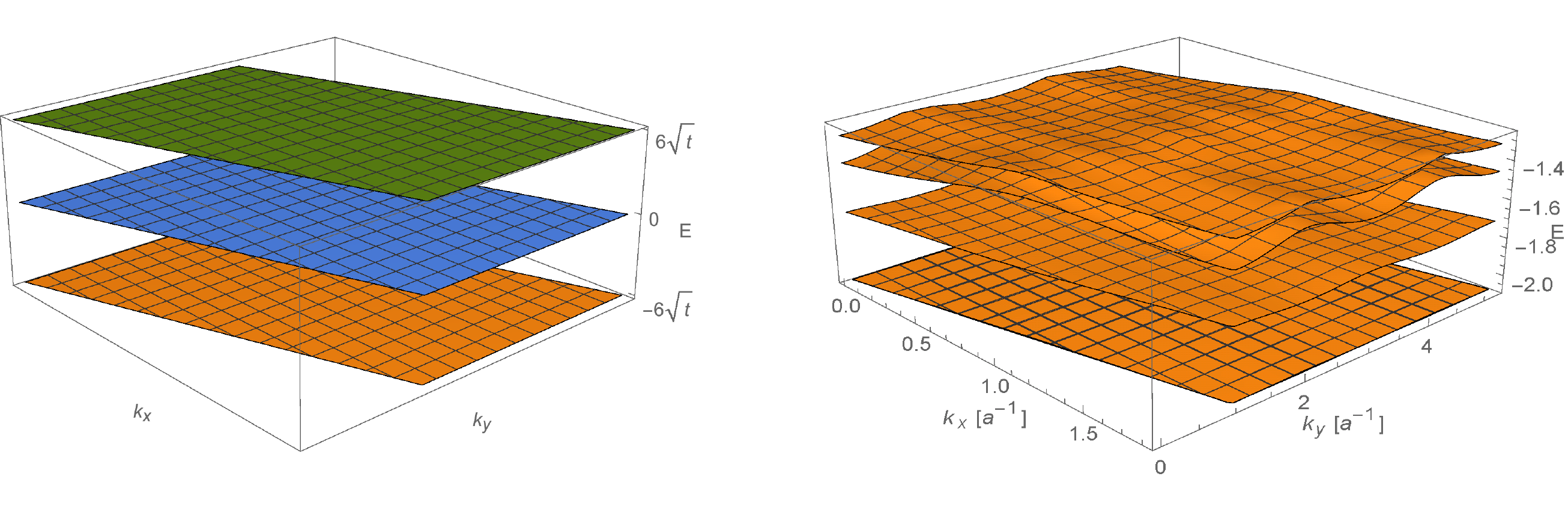} % combines previous two figures for better arrangement.
      \begin{picture}(0,0)
              \put(-420,115){(a)}
              \put(-210,115){(b)}
      \end{picture}
\caption{a) Spectrum of the fully frustrated dice-lattice model in the tight-binding limit, plotted over the first BZ. As the magnetic unit cell has six distinct sublattices, the model results in six bands that are pairwise degenerate with energies of $E=-\sqrt{6}t$,  $E=0$, and  $E=\sqrt{6}t$ for the three pairs of bands. b) Spectrum of the dice-lattice model with system parameters $\mathcal{V}=2 E_R$, and $-r=b=1/8$. The plot shows the lowest five bands, of which the lowest two energy bands are near-degenerate.}
\label{fig:DiceTightBinding}
\end{center}
\end{figure}
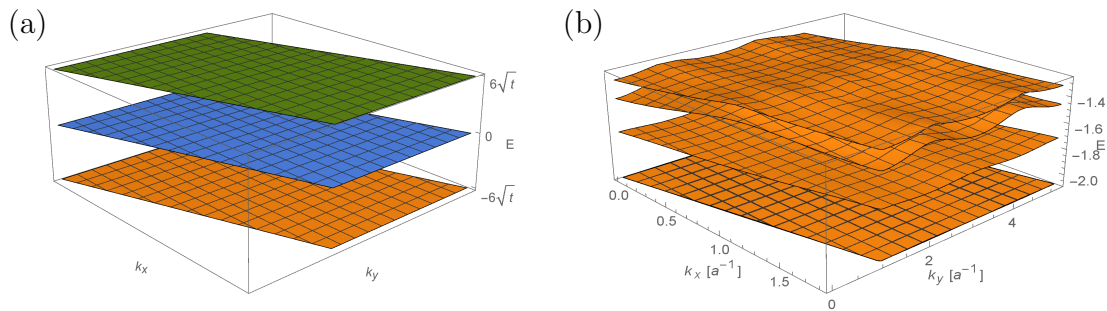

We proceed to discuss the spectrum, which provides an excellent approximation to the tight-binding version of the $\pi$-flux dice-lattice model. For reference, let us first review the spectrum in the tight-binding limit, shown in Fig.~\ref{fig:DiceTightBinding}a). Note that the tight-binding spectrum features only three distinct eigenvalues, each corresponding to a pair of degenerate bands all of which are time-reversal symmetric and have Chern number $C=0$. The overall count of six bands corresponds to the six lattice sites in the fundamental magnetic unit cell of the fully-frustrated dice-lattice. 

\begin{figure}[t]
\begin{center}
      \begin{picture}(0,0)
              \put(12,181){(a)}
              \put(120,181){(b)}
              \put(228,181){(c)}
              \put(342,181){(d)}
        \put(87,152){$E_0$}
        \put(196,152){$E_0$}
        \put(306,152){$E_0$}
        \put(417,152){$E_0$}
             \put(12,-14){(e)}
              \put(121,-14){(f)}
              \put(229,-14){(g)}
              \put(343,-14){(h)}
        \put(77,-40){$\log\left|\frac{\mathcal{B}}{a^2}\right|$}        
        \put(187,-40){$\log\left|\frac{\mathcal{B}}{a^2}\right|$}        
        \put(298,-40){$\log\left|\frac{\mathcal{B}}{a^2}\right|$}        
        \put(413,-40){$\log\left|\frac{\mathcal{B}}{a^2}\right|$}

      \end{picture}
\includegraphics[height=2.7in]{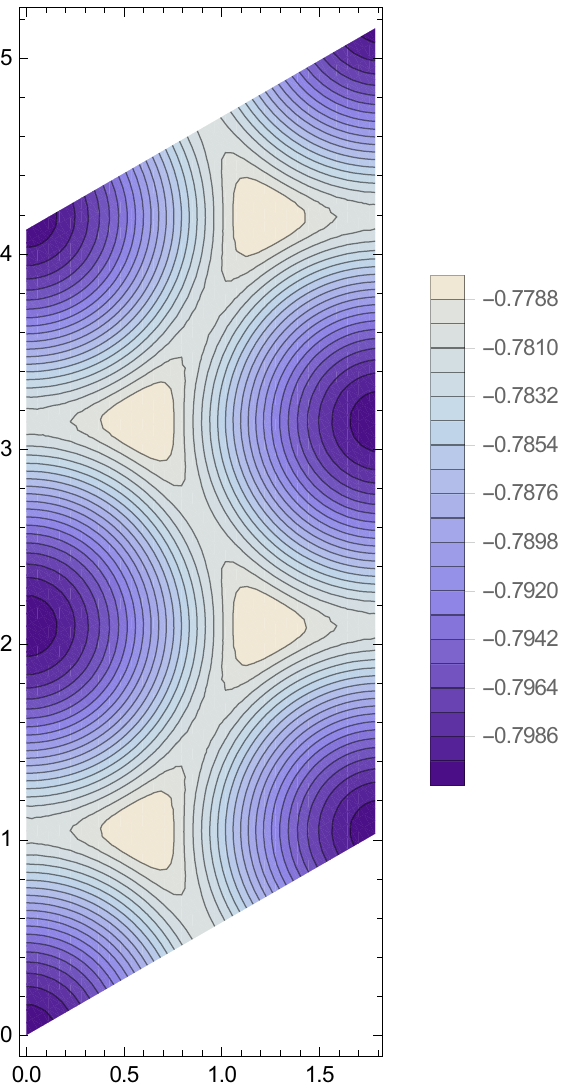}
\includegraphics[height=2.7in]{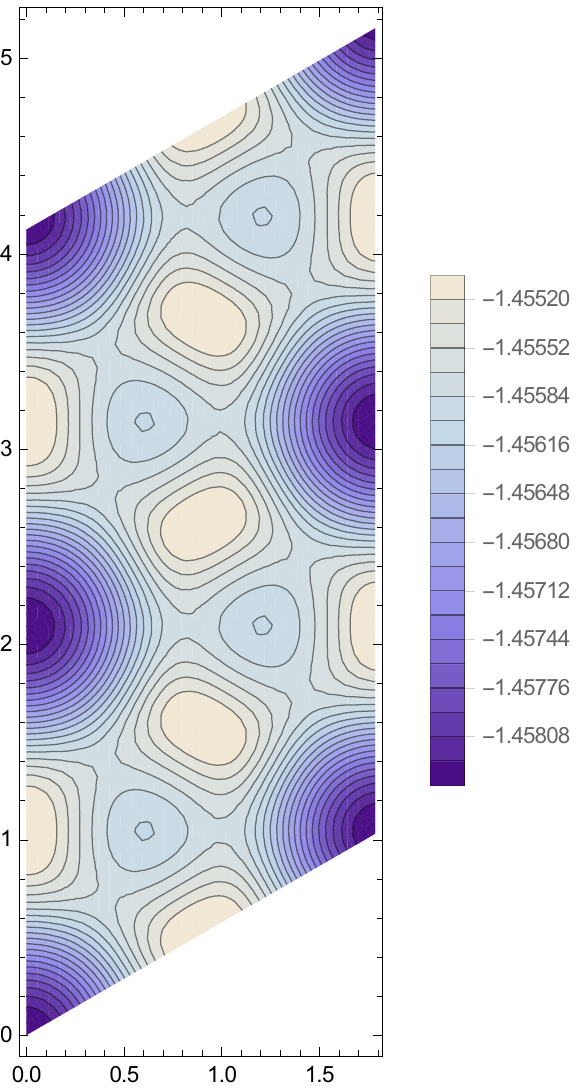}
\includegraphics[height=2.7in]{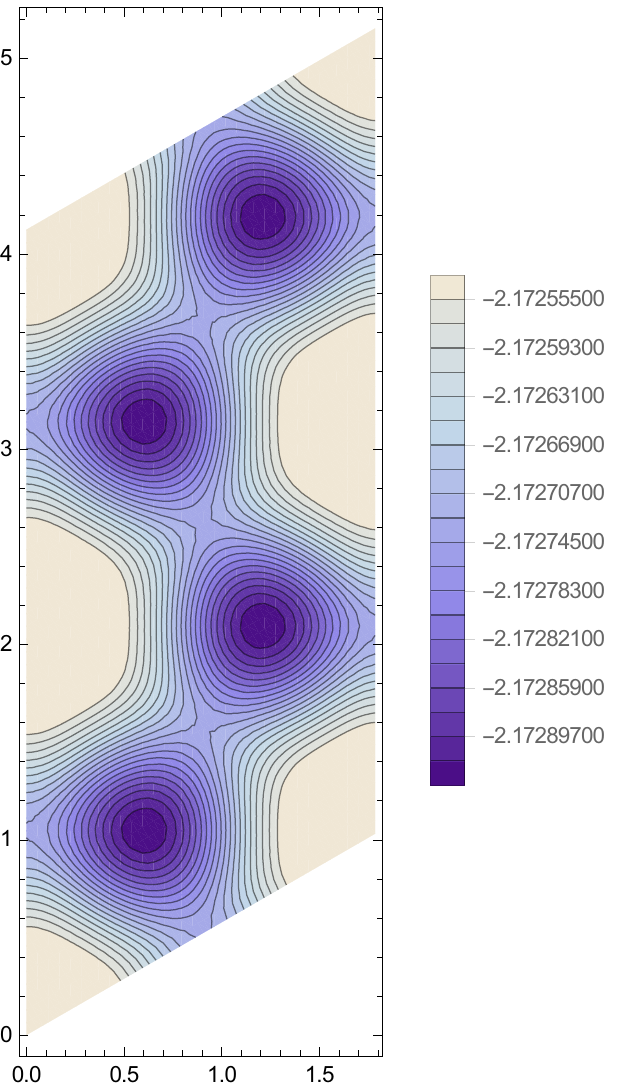}
\includegraphics[height=2.7in]{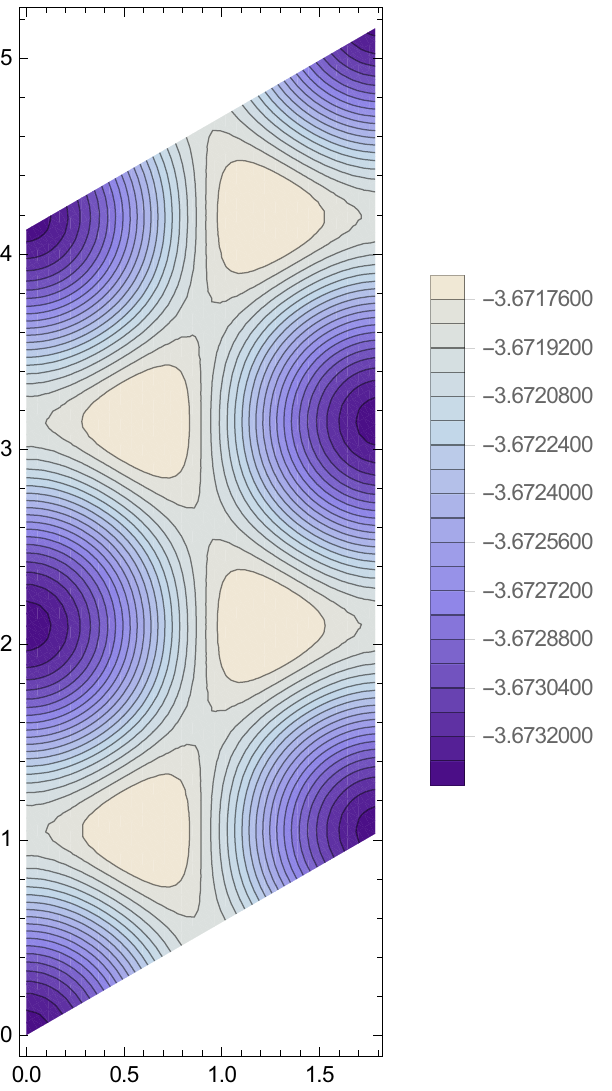} \\
\includegraphics[height=2.7in]{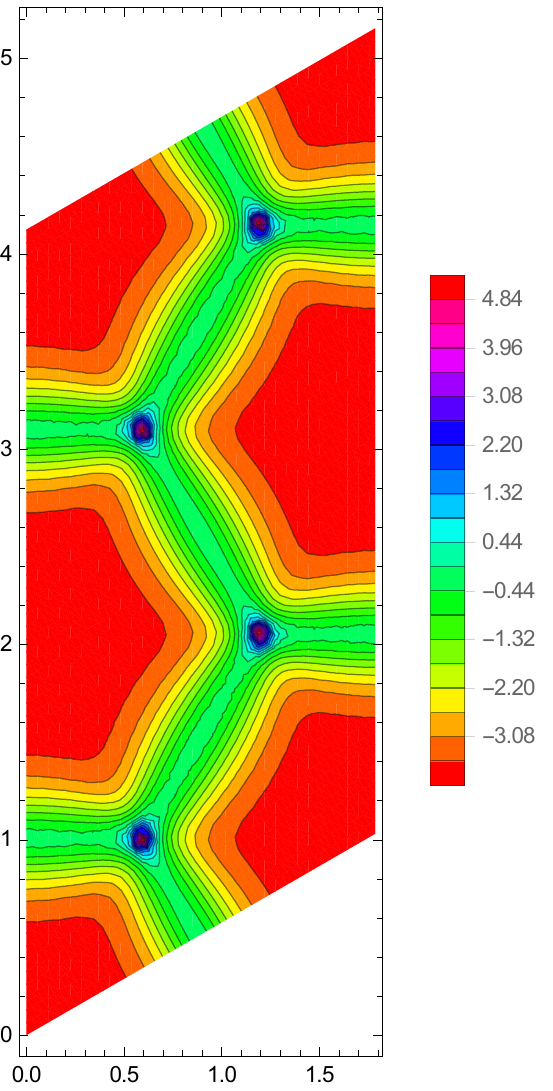}\hskip0.32cm
\includegraphics[height=2.7in]{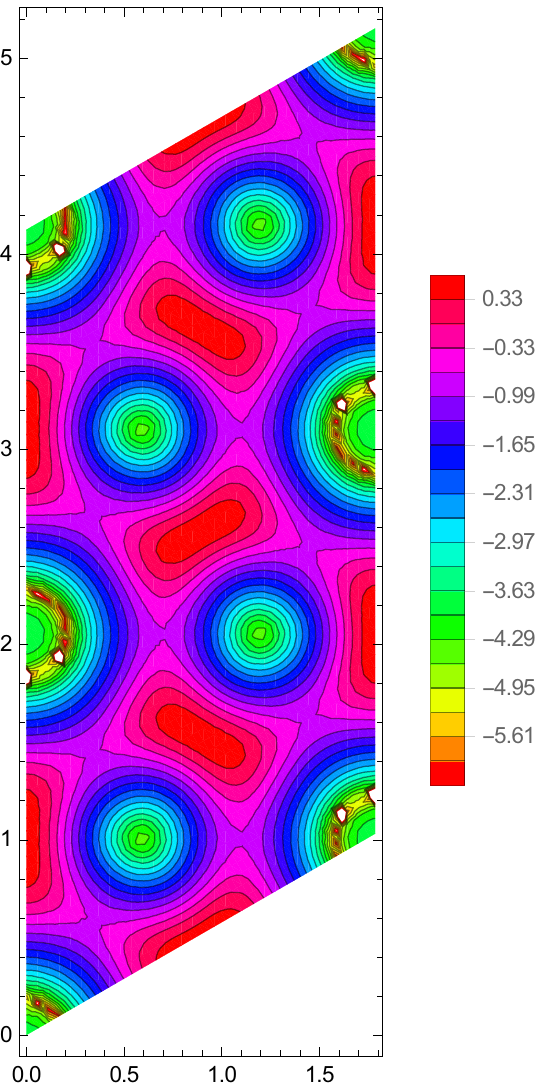}\hskip0.35cm
\includegraphics[height=2.7in]{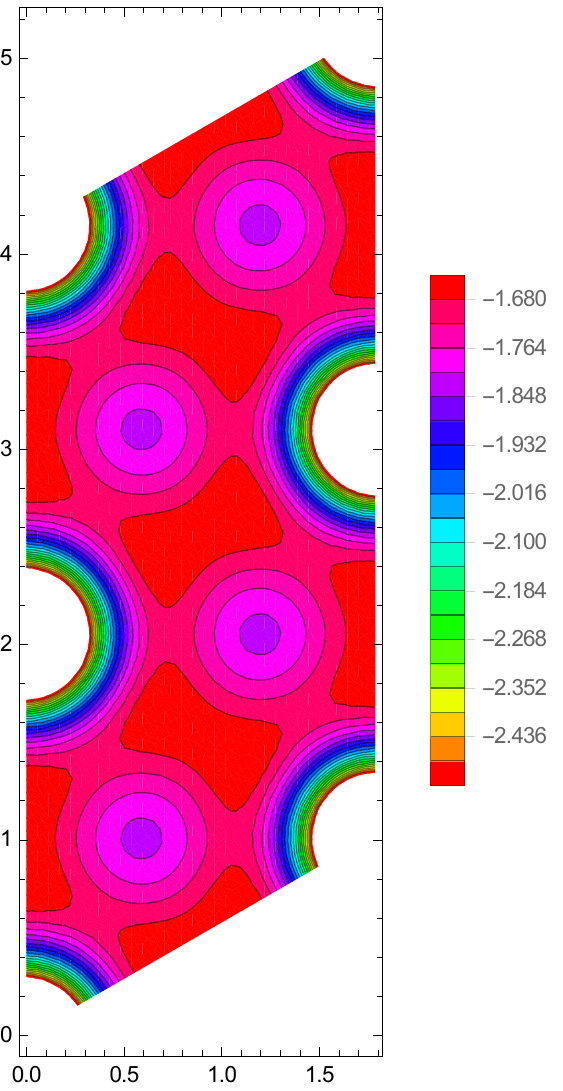}\hskip0.47cm
\includegraphics[height=2.7in]{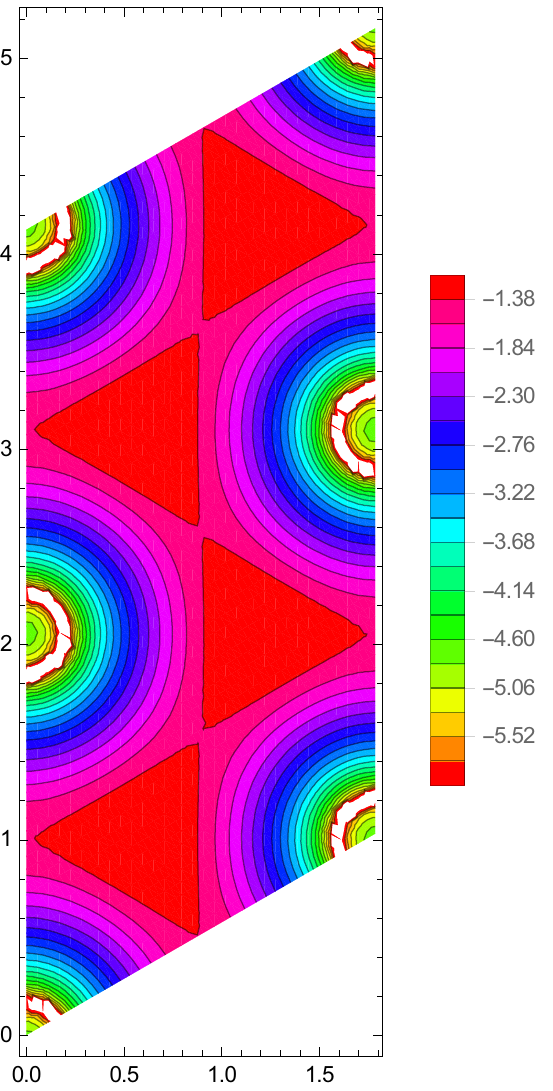}
\caption{Contour-plots of the properties of the lowest band in the dice-lattice model with system parameters $-r=b=1/6$ in the unfolded first Brillouin zone $[\cvec{g}_1,2\cvec{g}_2]$ for the energy (upper row) and logarithm of the magnitude of Berry curvature $\log |\mathcal{B}(k)a^{-2}|$ (bottom row). Values are shown for magnitudes of optical coupling $\mathcal{V}=E_R$ (panels a,e), $\mathcal{V}=1.5E_R$ (b,f), $\mathcal{V}=2E_R$ (c,g), and $\mathcal{V}=3E_R$ (d,h).}
\label{fig:DiceFluxLatticeContours}
\end{center}
\end{figure}

At intermediate depth of the optical lattice $\mathcal{V}\simeq E_R$,\footnote{Here, we define the recoil energy $E_R$ as in (\ref{eq:RecoilEnergy}), using the wave number $\kappa$ for the underlying triangular lattice as the reference. Although this is not the largest momentum transfer in the set up, it is the laser requiring the largest amplitude.} we find that the low-energy spectrum of our proposed dice flux lattice (\ref{eq:DiceFL}) correctly reproduces the qualitative features of the tight binding model. For $\mathcal{V}\simeq E_R$, this low-energy spectrum contains two near-degenerate bands that are well separated from higher bands. %Let us first focus on the nature of these lowest bands; we will return to discuss the fate of the higher tight-binding bands later.
These two lowest bands have a very small dispersion and have only a small residual splitting.  A typical spectrum, for $\mathcal{V}=2 E_R$, and $-r=b=1/8$ is shown in Fig.~\ref{fig:DiceTightBinding}b). %Fig.~\ref{fig:DiceFluxLatticeSpectrum_V2}. 
To display the residual dispersion of the lowest bands more clearly, we will analyse a series of contour-plots in Fig.~\ref{fig:DiceFluxLatticeContours}, below. For the parameters in Fig.~\ref{fig:DiceTightBinding}b), the dispersion of the two lowest bands is of the order of $0.04 E_R$. There is a small splitting to the second band (not shown), which has the inverse dispersion relative to that of the lowest band, i.e. its minima are found at the maxima of the lowest band and vice versa. With these parameters, the joint dispersion of these nearly degenerate bands is about 50 times smaller than the gap to higher excited bands.

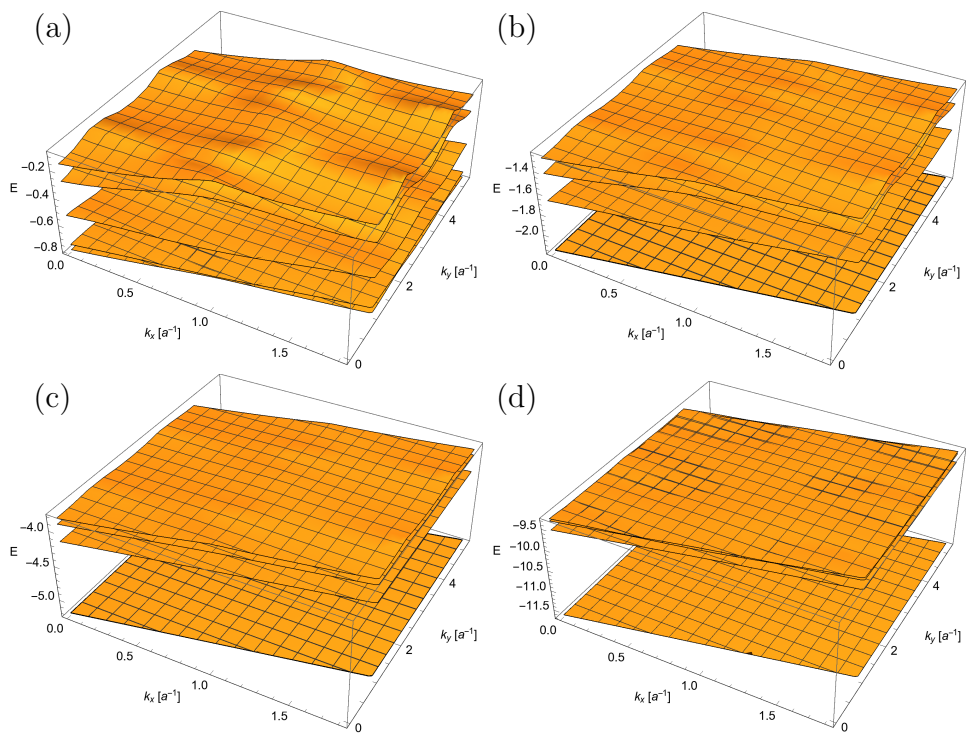
\begin{figure}[t]
\begin{center}
\includegraphics[width=0.4\columnwidth]{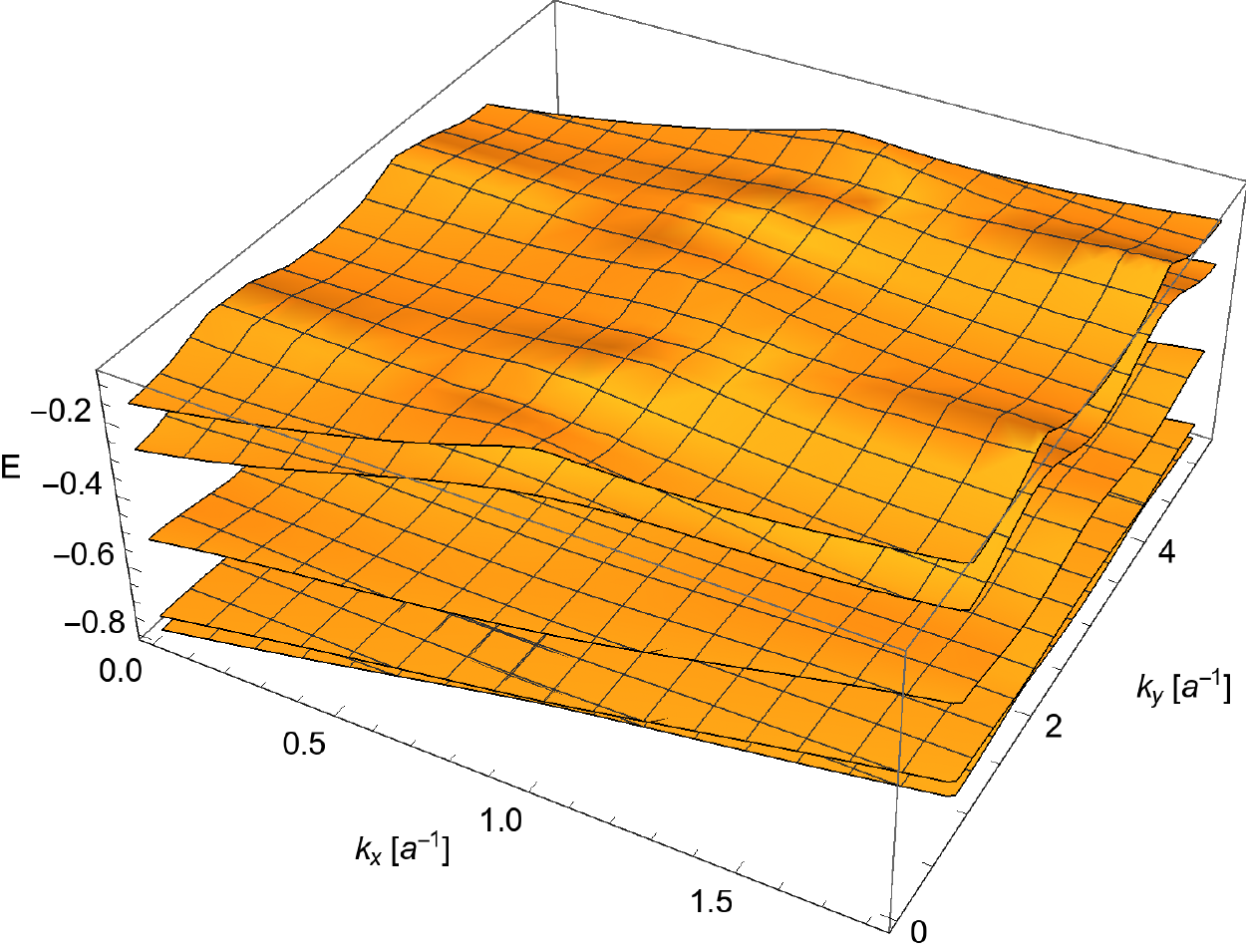}
\includegraphics[width=0.4\columnwidth]{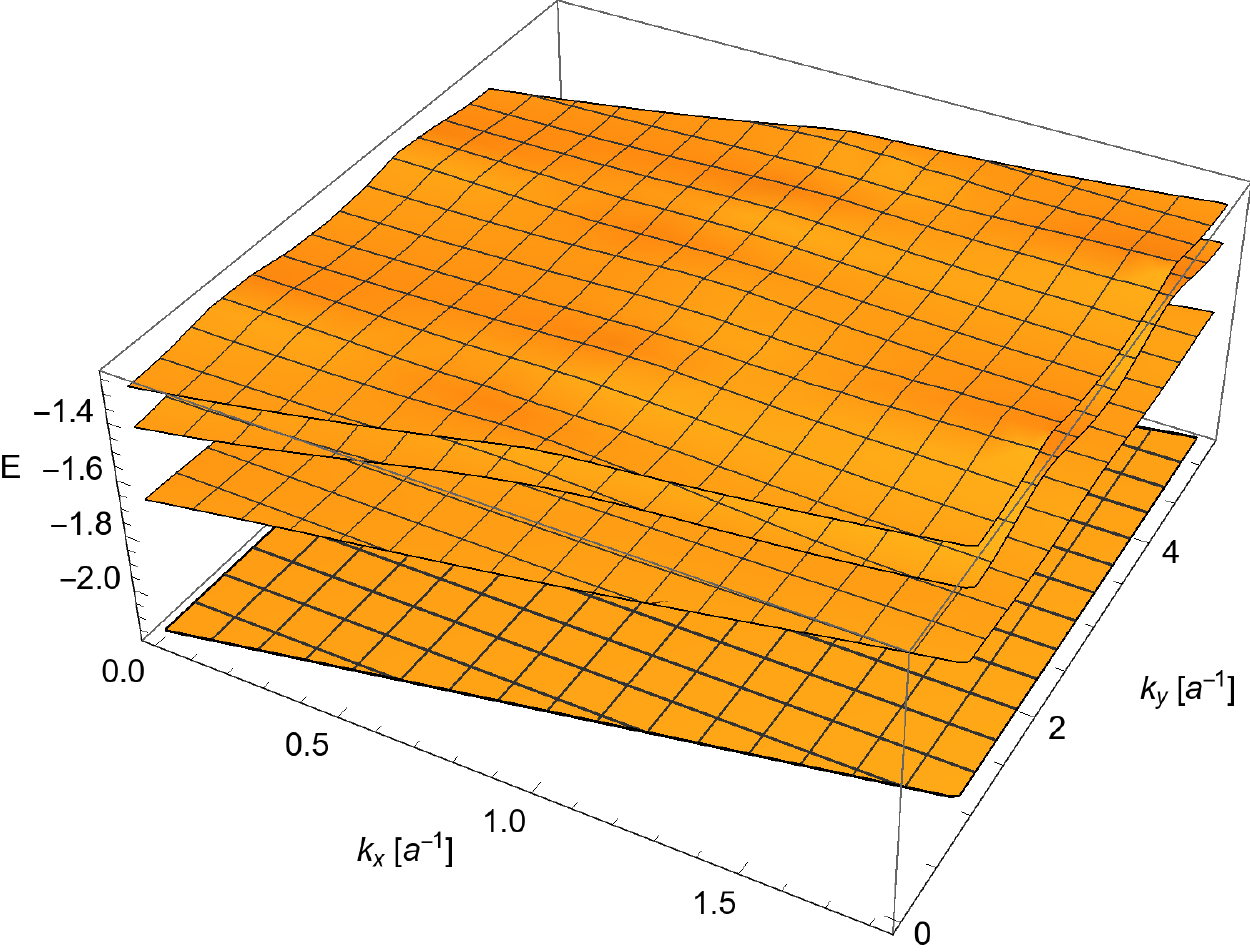}\\
\includegraphics[width=0.4\columnwidth]{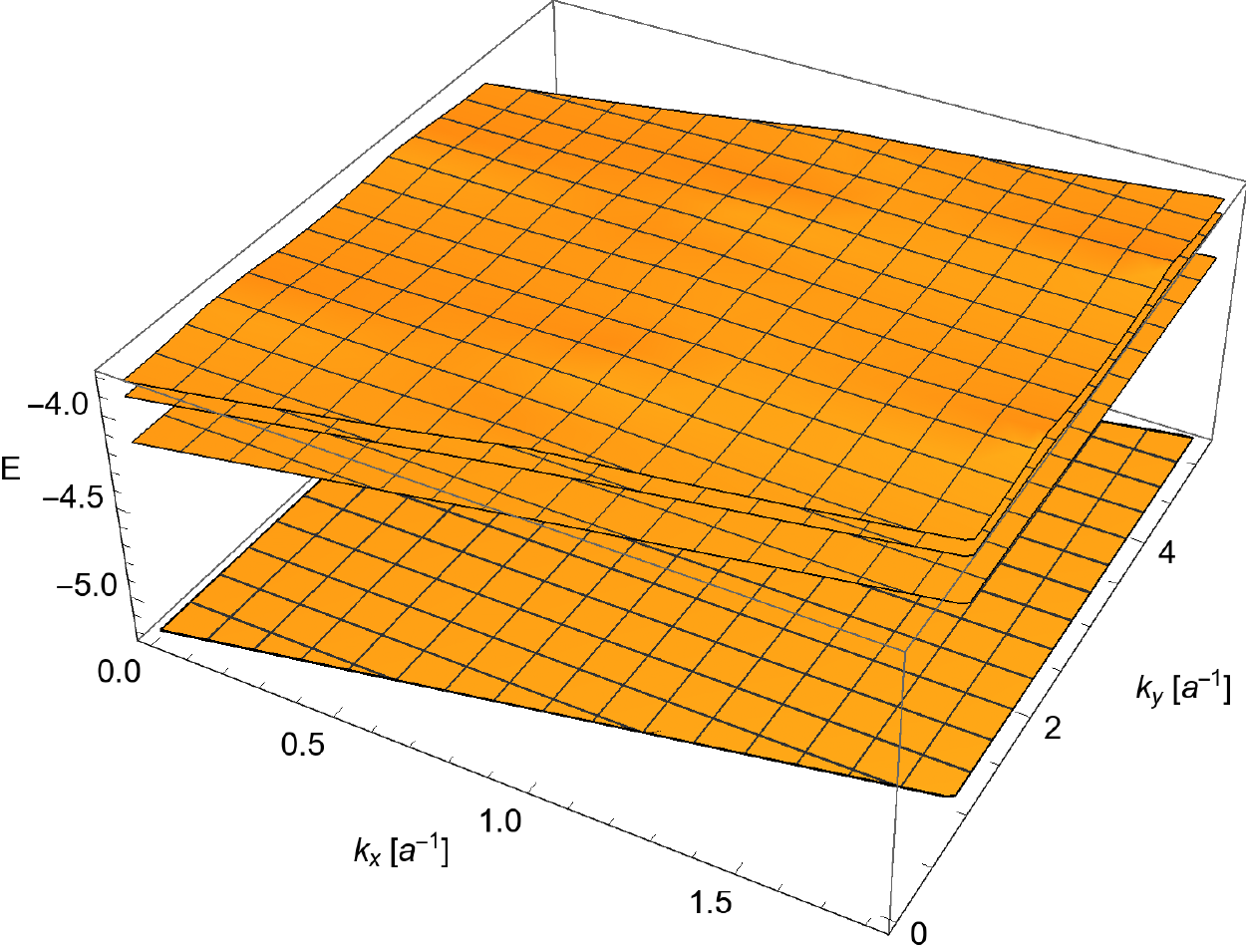}
\includegraphics[width=0.4\columnwidth]{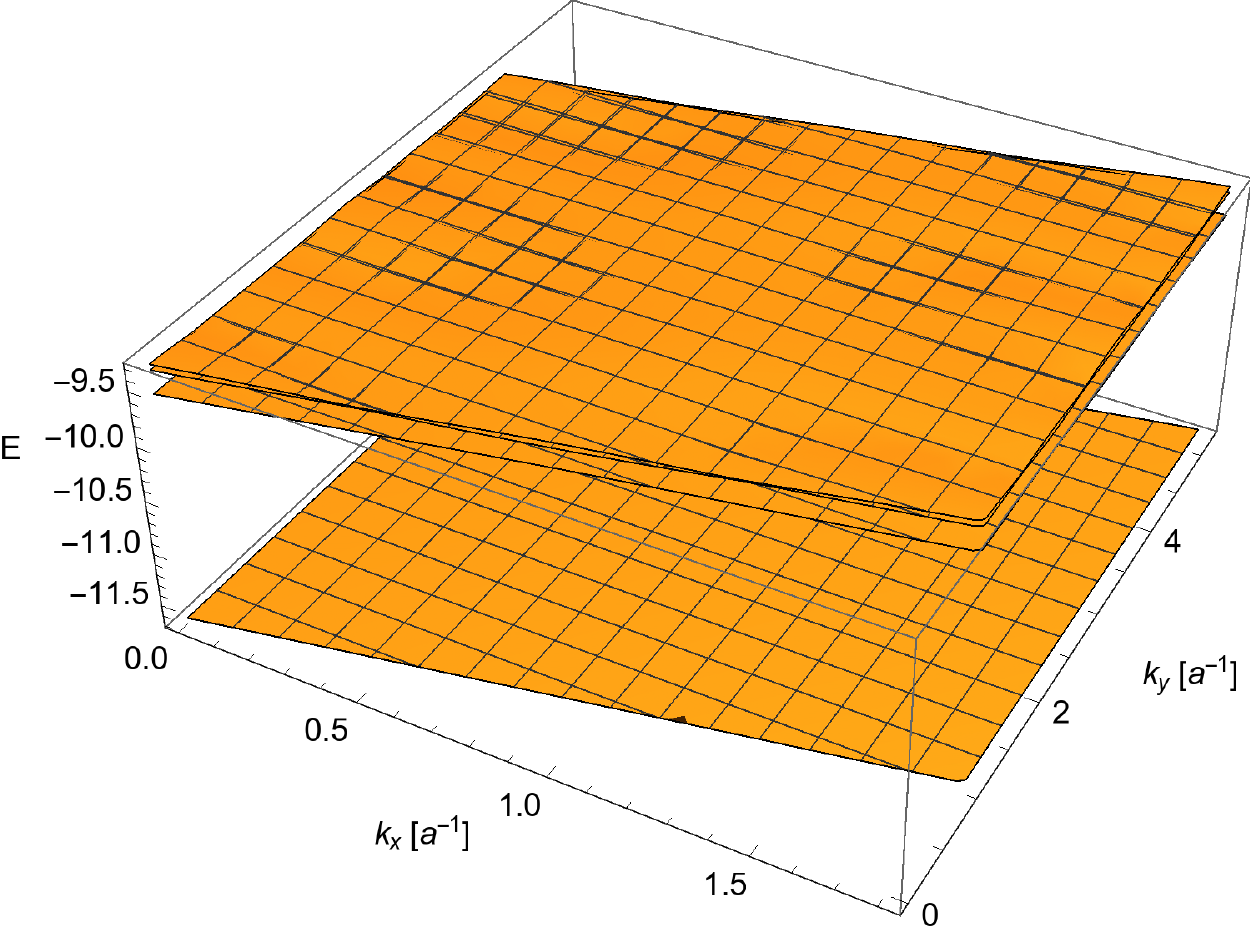}\\
      \begin{picture}(0,0)
              \put(-170,280){(a)}
              \put(5,280){(b)}
              \put(-170,140){(c)}
              \put(5,140){(d)}
      \end{picture}
\caption{Evolution of the spectrum of the dice-lattice model with system parameters as a function of the parameter $\mathcal{V}$, with fixed $-r=b=1/6$, shown within the enlarged Brillouin zone spanned by $[\cvec{g}_1,2\cvec{g}_2]$. The plots show the lowest five energy bands, of which the lowest two energy bands are near-degenerate. Values of $\mathcal{V}$ shown are (a) $\mathcal{V}=E_R$, (b) $\mathcal{V}=2E_R$, (c) $\mathcal{V}=4E_R$, and (d) $\mathcal{V}=8E_R$. Note how the gap above the pair of near-degenerate bands grows relative to the splitting of higher bands, as well as the overall increase in the magnitude of energy eigenvalues.
}
\label{fig:DiceFluxLatticeSpectrum_V4}
\end{center}
\end{figure}

It is instructive to analyze how the band dispersion evolves with the strength $\mathcal{V}$ of the optical coupling. A series of different spectra with values ranging from $\mathcal{V}=E_R$ to $\mathcal{V}=8E_R$ is shown in Fig.~\ref{fig:DiceFluxLatticeSpectrum_V4}, including the lowest five bands in each case. These data were obtained with a cut-off for momentum at $k\simeq 12 |\cvec{g}_i|$. It is clearly seen that the near-degeneracy of the lowest two bands is realised very well for all $\mathcal{V}\geq 2E_R$, while a small splitting is visible on the figure for $\mathcal{V}=E_R$. The higher ($n=3,4,5$) bands are not found to be degenerate. However, the gap above the near-degenerate ground state manifold is seen to increase with the optical coupling strength. Given these findings, we interpret the lowest bands as corresponding to the two degenerate lowest energy bands in the tight-binding limit, while the higher bands can be interpreted as arising from different local orbitals that can be formed within the minima of the optical potential. In the limit of $\mathcal{V}\to\infty$, we expect that the splitting to such orbitals would become large, and a low energy part corresponding to the single orbital physics may then emerge from the spectrum.

We now discuss the topological nature of the low-lying bands in the dice flux-lattice. The main qualitative difference of the intermediate-depth lattice with respect to the tight-binding model is the occurrence of weak tunnelling across the `forbidden' links of the underlying triangular flux lattice, which break time-reversal symmetry. To analyse this statement quantitatively, we calculate the Berry curvature $\mathcal{B}$ of our model by evaluating Wilson loops on a discretized grid of $k$-points within the Brillouin zone \cite{Fukui:2005ii}. We confirm that the Berry curvature is non-zero, and has opposite signs in the two low-lying bands. The distributions of the (log-)Berry curvature in the lowest band are shown as contour plots in the lower row of Fig.~\ref{fig:DiceFluxLatticeContours} for a range of optical coupling strengths, while the upper row shows the corresponding band dispersions. Note that there are extended regions where the curvature $\mathcal{B}$ is small, while maxima are relatively localised. For example, at $\mathcal{V}=E_R$, typical values are $\mathcal{B}\simeq 0.05 a^{2}$ (to be compared to an average of $\bar \mathcal{B} = 9/\pi \mathcal{C}a^2 \simeq 2.86 C a^2$ for a Chern number $\mathcal{C}$ band with homogenous Berry curvature of the given Brillouin zone area). At the location of the maxima of the band dispersion, which can be seen as avoided crossings with the next higher band, $\mathcal{B}$ is strongly peaked and as a result, the Chern number $\mathcal{C}$ of the band is non-zero. Depending on the specific parameters we have found either $|\mathcal{C}|=1$, or $|\mathcal{C}|=3$. In both cases, the cumulative Chern number of the two lowest bands is zero. 

The different panels of Fig.~\ref{fig:DiceFluxLatticeContours} show the evolution of the band dispersion with increasing optical coupling, which reveals a change of the location of minima in the dispersion, and correspondingly for the Berry curvature. Note also how the flatness of the bands improves as we go to stronger coupling. Extended regions of low Berry curvature are also found at the highest value we show.

It would be interesting to study how the many-body spectrum is affected by this finite but oppositely oriented Berry curvature in the lowest two bands. We expect that as long as the interaction energy is larger than the residual splitting between the two lowest bands, the system likely behaves in a qualitatively similar fashion as the time-reversal invariant system in the tight binding limit \cite{2012PhRvL.108d5306M}. A detailed analysis of this physics will be the subject of a future study. In the sense that the perturbation of the bands away from the time-reversal symmetric case is caused by small hopping elements on suppressed bonds, we can consider the time-reversal symmetry breaking of our optical dice flux lattice to be `weak'.

\section{Realizing the Fully Frustrated Dice Lattice in a Tight-Binding Approach}
\label{sec:DiceTightBinding}

An alternative realisation of the dice lattice pierced by $\pi$-flux per plaquette can be realised in a pure tight-binding philosophy. Let us discuss in  detail the set-up for alkaline earth atoms [e.g., ytterbium (Yb)] atoms trapped in an optical lattice at the anti-magic wavelength \cite{2012PhRvL.108d5306M}. 
In our approach, we closely follow the proposal for a square optical lattice using anti-magic trapping \cite{Gerbier:2010ho}.
The possibility for this construction arises as the two internal states ($^1S_0$ and $^3P_0$) of Yb have polarisability $\alpha$ of opposite signs for wavelengths $\lambda \gtrsim 960$nm, so they are trapped at the points of maximum or minimum laser intensity, respectively \cite{Gerbier:2010ho}. At the anti-magic wavelength $\lambda^*\simeq 1120$nm, the absolute values of the polarisability are of equal magnitude. This is crucial for the square lattice geometry. For our purposes, it may actually be more useful to choose a wavelength at which the polarisability is stronger in magnitude for one of the two (pseudo-)spin states: the dice lattice geometry results from a triangular optical lattice formed by three self-reflected laser beams propagating with wave vectors arranged at relative angles of $2\pi/3$ with respect to each other. These beams should be mutually incoherent, so the total intensity is the sum of individual intensities. The mirrors used to self-reflect these beams need to be stabilised.\footnote{Alternatively, a suitable triangular lattice potential can be generated by three running beams with relative phase coherence. However, these would additionally have to be phase stabilised to prevent this triangular lattice from drifting relative to the 4th standing wave, laser $S_4$, discussed below.} One species of atoms ($^1S_0$) is then trapped at the maxima of the intensity (which are steep), while the excited $^3P_0$ state is trapped at the minima (which are more shallow). Hence, it is favourable that the polarisability is larger for the excited state, implying use of a wavelength $\lambda_0 \equiv 2\pi/k_0 > \lambda^*$, i.e.~using wavelengths in the far infrared (given that the polarisablity of the excited state grows more rapidly with $\lambda$ near the anti-magic wavelength, or $d\alpha(^3P_0)/d\lambda|_{\lambda^*} > d\alpha(^1S_0)/d\lambda|_{\lambda^*}$).

\begin{figure}[t]
\begin{center}
\includegraphics[width=0.99\columnwidth]{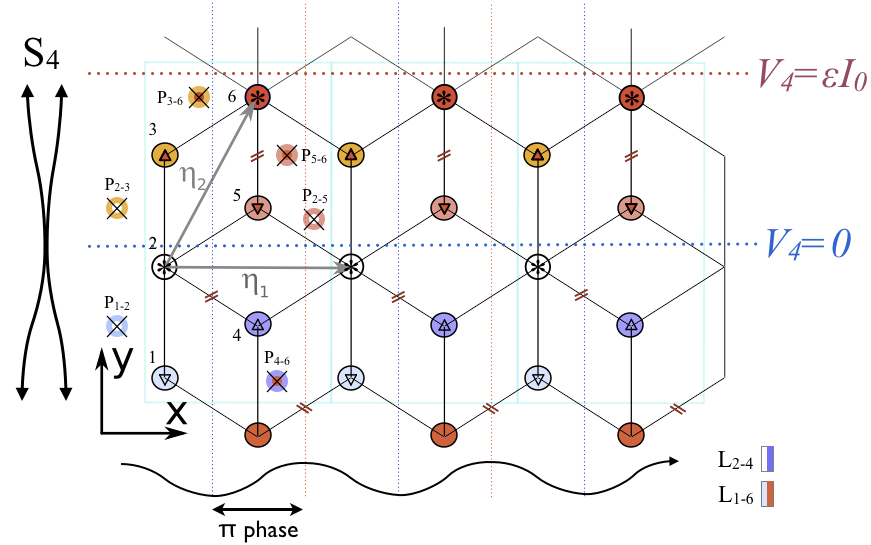}
\caption{Illustration of the rectangular magnetic unit cell with six inequivalent sites numbered $1$ to $6$. The drawing includes three magnetic unit cells, delineated by light solid lines. Links indicate the connectivity of the lattice, corresponding to hopping with amplitude $t$. Three links in the magnetic unit cell are special and need to be chosen with negative hopping $-t$ (shown with two hashes). One laser, $S_4$ is required to establish the magnetic unit cell. Hopping between the six energetically inequivalent sites of the magnetic Brillouin zone are driven by lasers as indicated. Hopping-inducing lasers propagating perpendicular to the plane are labelled $P_{i-j}$ and drive transitions between sites $i$ and $j$ (shown as circles with crosses). The last two lasers $L_{i-j}$ propagate with a non-zero in-plane momentum along the $x$-axis such as to induce two distinct transitions within each magnetic unit cell, and with the relatively opposite sign.}
\label{fig:OpticalDiceLatticeTransitions}
\end{center}
\end{figure}

In our set-up, all neighbouring sites are occupied by atoms of different internal states. Consequently, spontaneous tunnelling processes can be neglected, and all dynamics in this lattice is driven by via laser-assisted hopping \cite{Ruostekoski:2002fs,Jaksch:2003ud}.
Simultaneously, this coupling enables one to imprint phases onto the hopping matrix elements \cite{Gerbier:2010ho}.
Let us now explain how to achieve phases that yield the target flux density of $n_\phi=1/2$. For the fully frustrated dice lattice, the magnetic unit cell contains six inequivalent atoms \cite{Vidal:1998gx,2012PhRvL.108d5306M}, chosen here as a rectangular cell spanned by vectors $\cvec{v}_1 = (\sqrt{3}a,0)^t$ and $\cvec{v}_2 = (0,3a)^t$, as indicated by the different colouring of inequivalent lattice sites in Fig.~\ref{fig:OpticalDiceLatticeTransitions}. However, the (scalar) triangular optical lattice described in the preceding paragraph distinguishes only two types of sites. We propose to break this symmetry by shining one additional self-reflected laser-beam, $S_4$, onto the system: this beam serves to break down the internal mirror-symmetry of the triangular lattice unit cell to the desired periodicity. In our set-up, $S_4$ has the same frequency/wavelength as the triangular optical lattice. However, its in-plane wavelength is enlarged to $\lambda_4^\parallel=\lambda_0/\sin(\theta)$ by projecting this laser onto the system at a tilt angle $\theta$ with respect to the $z$-axis of the plane. We tilt the laser towards the $y$-direction and require the potential to repeat on the scale of the magnetic unit cell, i.e., $|\cvec{v}_2|=\lambda_4^\parallel/2$. By geometry, we must therefore choose the angle $\theta = \arcsin(1/2)=\pi/6$. Note the position of this laser potential ($S_4$) needs to maintain a fixed spatial position relative to the lasers defining the optical lattice, as fluctuations would shift the superlattice potential relative to the triangular lattice potential, and would alter the relative magnitudes of site energies. However, these energies need to be  precisely defined, so that coupling lasers can satisfy the resonance condition and match the binding energy differences for the links on which they induce hopping processes. Note that a different wave length laser could also be chosen.

To be explicit, let us write the required laser potentials. A bare triangular optical lattice of lattice constant $a$ is created by the wavelength $\lambda_0=3a$ of the trapping beams:
\begin{eqnarray}
\label{eq:TriangularTightBinding}
V_\mathrm{tri}(\br) = I_0 \sum_i \sin^2 \left(\frac{2\pi}{3a}\cvec{\kappa}_i \cdot \br\right), 
\end{eqnarray}
with the unit lattice directions $\cvec{\kappa}_1 = (0,1,0)^t$, $\cvec{\kappa}_2=(-\sqrt{3}/2,1/2,0)^t$, and $\cvec{\kappa}_3=(\sqrt{3}/2,1/2,0)^t$. The additional self-reflected laser, $S_4$, propagates along the direction $\cvec{n}_d = (0,\sin\theta,\cos\theta)^t$, adding an (in-plane) intensity distribution of
\begin{eqnarray}
\label{eq:DetuningTightBinding}
V_4(\br) = \epsilon I_0 \sin^2 \left(\frac{2\pi}{3a}\vec n_d \cdot \br+\delta \right) \equiv  \epsilon I_0 \sin^2 \left(\frac{2\pi}{6a} y+\delta \right)
\end{eqnarray}
Here, we need to choose a small offset of the phase $\delta$ such that the maximum of intensity of the additional laser does not align with any high-symmetry point in the magnetic unit cell, and the intensity of the inversion symmetry breaking laser $S_4$ is reduced with respect to the other lasers by a suitable small factor $\epsilon$, e.g., we can choose number of the order $\delta\simeq 2\pi/10$ and $\epsilon\simeq 0.05$.

\begin{figure}[t]
\begin{center}
\includegraphics[width=0.6\columnwidth]{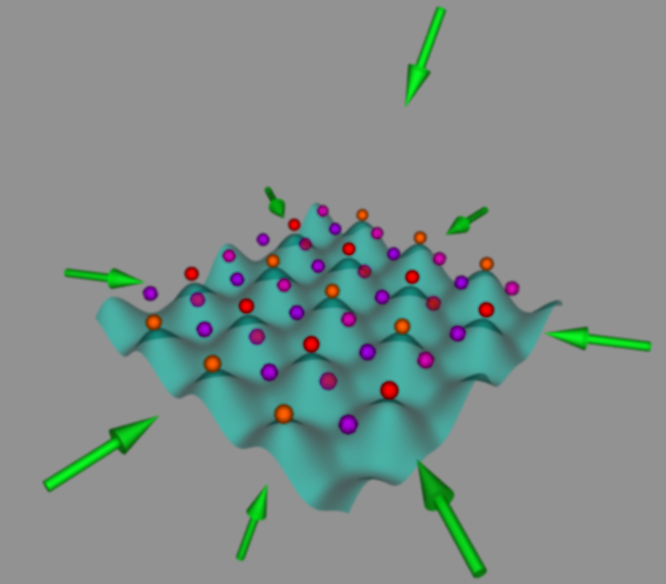}
\caption{Set up of an optical dice lattice using an anti-magic optical lattice with laser-induced hopping: an underlying triangular lattice is created by retro-reflected standing wave lasers in plane. The symmetry of the magnetic unit cell is created by an additional standing wave laser $S_4$ directed at an angle to the plane. Eight coupling lasers complete the set-up and drive transitions between sites of different energy, as shown in Fig.~\ref{fig:OpticalDiceLatticeTransitions} and discussed in the main text.}
\label{fig:RamanOpticalDiceLatticeSetUp}
\end{center}
\end{figure}

A three-dimensional view of the overall set-up is given in Fig.~\ref{fig:RamanOpticalDiceLatticeSetUp}. In the resulting potential $V_\mathrm{tot}(\br)=V_\mathrm{tri}(\br)+V_4(\br)$, the six sublattices of the desired magnetic unit cell are all distinguished energetically, i.e.~their energies being detuned with respect to the triangular lattice by distinct amounts $\delta\epsilon_i$, $i=1,\ldots,6$. 

The set-up is completed by a total of eight coupling lasers driving the respective transitions between these sites. All of these lasers are propagating waves. Six of them are directed onto the system in the direction perpendicular to the lattice-plane. We denote these lasers as $P_{i-j}$, indicating the two lattice sites $i$, $j$ between which they induce a resonant transition. The six required lasers are $P_{1-2}$,  $P_{2-3}$, $P_{2-5}$, $P_{3-6}$, $P_{4-6}$, and $P_{5-6}$, which require frequencies $\hbar \omega_{i-j}=\hbar\omega^0_i +\delta\epsilon_i - (\hbar \omega^0_j +\delta\epsilon_j)$, and $\hbar\omega^0_i$ denotes the unperturbed energy of the internal state trapped at site $i$. Note each laser drives a transition between two neighbouring sites where atoms are in their ground / excited state, respectively. See also Fig.~\ref{fig:OpticalDiceLatticeTransitions} for an illustration. Four of these six lasers drive a transition on a single link in the unit cell. However, the two lasers $P_{2-5}$, $P_{3-6}$ connect the sixfold connected sites `$2$' and `$6$' to two neighbours with identical energy, located in the same and an adjacent unit cell, respectively. Due to the perpendicular direction of the lasers, these transitions are driven in phase, so the hopping elements have the same sign. All but one of the lasers $P_{i-j}$ need to be in phase with each other, while $P_{5-6}$ requires a phase-shift of $\pi$ relative to the others. A definite phase relationship between these lasers of different frequencies can be achieved by deriving them from a single light source, and detuning their frequency using an acousto-optic modulator. The remaining coupling two lasers, which we call $L_{1-6}$ and $L_{2-4}$ are special in that they are required to drive two transitions (like $P_{5-6}$), but now with a relative phase of $\pi$ between these two couplings. This relative phase is realised by virtue of an in-plane component of the respective wave-vectors. Specifically, we choose the in-plane component of their respective wave-vectors $\mathbf{k}$ along the $x$-axis such that $\mathbf{k}\cdot (\sqrt{3}a/2,0,0)^t\equiv \pi$. Again, this wave-vector can be realized by a suitable inclination of the laser beams with respect to the plane.

This concludes our discussion of the detailed set-up for a tight-binding version of fully frustrated dice lattice. Let us briefly compare this construction to the optical dice flux lattice discussed in section \ref{sec:DiceQuantitative}. Firstly, we note that the tight-binding construction is explicitly time-reversal invariant, if all relative phases are set to match the values $0$ or $\pi$. Although there may be small perturbations to the ideal dice-lattice model from spontaneous tunnelling processes between neighbouring three-fold sites such as sites $1$ and $4$, such processes also have real hopping elements.

The practical realisation of both schemes poses similar challenges, notably the requirement to generate superlattice potentials whose relative position must be stabilised relative to an underlying lattice. This is difficult, but has already been achieved \cite{Jo:2012br}. However, fluctuations of the geometry will affect the two proposals rather differently. In the optical flux lattice set-up, the superlattice acts to suppress tunnelling by creating local maxima in the potential. This suppression will be relatively insensitive to the precise location of potential maxima, as long as they are located within the relevant bonds of the lattice. By contrast, the tight-binding approach requires the superlattice to define relative energies of lattice orbitals, and transitions between them are driven resonantly. Hence, a rather fine control of the stability is required to ensure that all coupling lasers remain on resonance for their respective bonds.

\section{Conclusions}
\label{sec:Conclusions}

We have introduced a new method for constructing optical flux lattices with complex geometries by combining a simple optical flux lattice with additional scalar potentials. To demonstrate the potential of our proposal, we have explored the optical dice flux lattice as an example geometry in which bonds were eliminated from an underlying triangular lattice. Our model yields flat bands that are a particularly interesting playground for studying interaction-driven phases of matter \cite{2012PhRvL.108d5306M}, and can realise a flatness parameter of fifty even for weak optical coupling. The optical flux-lattice approach results in interesting additional features with respect to a pure tight-binding description of the dice-lattice model. At intermediate lattice depths, the model weakly breaks time-reversal symmetry in the following sense: instead of degenerate pairs of time-reversal symmetric bands, the approach produces time-reversal pairs of bands whose degeneracies are only weakly split.

The proposed realisation of an optical dice flux lattice is realistically achievable in the near future, as it combines several elements which are already part of the current state of the art. The kagome lattice realised in the group of Stamper-Kurn successfully demonstrates the phase-stabilised superposition of two lattices with distinct wavelengths \cite{Jo:2012br}. Our set-up requires the additional superposition of a triangular optical flux lattice. While work on the first realisation of such systems under way, we would like to underline that related schemes for synthetic gauge fields have already been successful \cite{Aidelsburger:2011hl, 2010PhRvA..82f3625M}, and related schemes for emulating spin-orbit coupling in 2D systems have also been implemented \cite{Wu:2016kv, Huang:2016kf, Sun:2017wa}.
 
We have also introduced a proposal for a tight-binding scheme which is closer to the existing technology of the aforementioned experiments. Here, challenges rely on fine-tuning energies and maintaining the relative superlattice position with high accuracy. This kind of set-up requires one-by-one engineering of laser-induced hopping between sites in the unit cell, so its complexity grows with the unit cell size.

By contrast, one of the inherent features of the flux-lattice schemes is their tuneability. Explorations of scalar optical lattices have already shown that a multitude of different band-structures can be realised in the same experiment \cite{Jo:2012br,Tarruell:2012db}.
Hence, one interesting direction for further study is the question of how the lattice geometry is altered when moving the scalar lattices with respect to the underlying optical flux lattice. 

\ack
G.M. acknowledges support from the Leverhulme Trust under Grant No.~ECF-2011-565, from the Newton Trust, and from the Royal Society under grant UF120157. N.R.C acknowledges support by EPSRC Grant No.~EP/K030094/1. Statement of compliance with EPSRC policy framework on research data: All data accompanying this publication are directly available within the publication.

\section*{References}

\bibliographystyle{iopart-num}

\bibliography{dice-lattice-realizations}

\providecommand{\newblock}{}
\begin{thebibliography}{10}
\expandafter\ifx\csname url\endcsname\relax
  \def\url#1{{\tt #1}}\fi
\expandafter\ifx\csname urlprefix\endcsname\relax\def\urlprefix{URL }\fi
\providecommand{\eprint}[2][]{\url{#2}}

\bibitem{Cooper:2008hx}
Cooper N~R 2008 {\em Advances in Physics\/} {\bf 57} 539--616

\bibitem{Cooper:2013jg}
Cooper N~R and Dalibard J 2013 {\em Phys. Rev. Lett.\/} {\bf 110} 185301

\bibitem{Goldman:2014bvb}
Goldman N, Juzeliunas G, {\"O}hberg P and Spielman I~B 2014 {\em Rep. Prog.
  Phys.\/} {\bf 77} 126401

\bibitem{Goldman:2016fa}
Goldman N, Budich J~C and Zoller P 2016 {\em Nat Phys\/} {\bf 12} 639--645

\bibitem{Gross:2017do}
Gross C and Bloch I 2017 {\em Science\/} {\bf 357} 995--1001

\bibitem{AboShaeer:2001go}
Abo-Shaeer J~R, Raman C, Vogels J~M and Ketterle W 2001 {\em Science\/} {\bf
  292} 476--479

\bibitem{Cooper:1999bz}
Cooper N~R and Wilkin N 1999 {\em Phys. Rev. B\/} {\bf 60} R16279--R16282

\bibitem{Cooper:2001gy}
Cooper N~R, Wilkin N and Gunn J 2001 {\em Phys. Rev. Lett.\/} {\bf 87} 120405

\bibitem{Sorensen:2005bt}
S{\o}rensen A~S, Demler E and Lukin M~D 2005 {\em Phys. Rev. Lett.\/} {\bf 94}
  086803

\bibitem{Kol:1993wv}
Kol A and Read N 1993 {\em Phys. Rev. B\/} {\bf 48} 8890--8898

\bibitem{Palmer:2006km}
Palmer R~N and Jaksch D 2006 {\em Phys. Rev. Lett.\/} {\bf 96} 180407

\bibitem{2009PhRvL.103j5303M}
M{\"o}ller G and Cooper N~R 2009 {\em Phys. Rev. Lett.\/} {\bf 103} 105303

\bibitem{2012PhRvL.108y6809H}
Hormozi L, M{\"o}ller G and Simon S~H 2012 {\em Phys. Rev. Lett.\/} {\bf 108}
  256809

\bibitem{Moller:2015kg}
M{\"o}ller G and Cooper N~R 2015 {\em Phys. Rev. Lett.\/} {\bf 115} 126401

\bibitem{Andrews:2018il}
Andrews B and M{\"o}ller G 2018 {\em Phys. Rev. B\/} {\bf 97} 035159

\bibitem{Fetter:2009fh}
Fetter A 2009 {\em Rev. Mod. Phys.\/} {\bf 81} 647--691

\bibitem{Jaksch:2003ud}
Jaksch D and Zoller P 2003 {\em New J. Phys.\/} {\bf 5} 56--56

\bibitem{Mueller:2004hc}
Mueller E~J 2004 {\em Phys. Rev. A\/} {\bf 70} 041603

\bibitem{Eckardt:2005bq}
Eckardt A, Weiss C and Holthaus M 2005 {\em Phys. Rev. Lett.\/} {\bf 95} 260404

\bibitem{Lin:2009us}
Lin Y, Compton R, Jimenez-Garcia K, Porto J and Spielman I~B 2009 {\em
  Nature\/} {\bf 462} 628--632

\bibitem{Gerbier:2010ho}
Gerbier F and Dalibard J 2010 {\em New J. Phys.\/} {\bf 12} 033007

\bibitem{Dalibard:2011gg}
Dalibard J, Gerbier F, Juzeli{\=u}nas G and {\"O}hberg P 2011 {\em Rev. Mod.
  Phys.\/} {\bf 83} 1523--1543

\bibitem{Cooper:2011iv}
Cooper N~R 2011 {\em Phys. Rev. Lett.\/} {\bf 106} 175301

\bibitem{Thouless:1982kq}
Thouless D~J, Kohmoto M, Nightingale M~P and den Nijs M 1982 {\em Phys. Rev.
  Lett.\/} {\bf 49} 405--408

\bibitem{Haldane:1988gh}
Haldane F~D~M 1988 {\em Phys. Rev. Lett.\/} {\bf 61} 2015--2018

\bibitem{Kane:2005hl}
Kane C~L and Mele E~J 2005 {\em Phys. Rev. Lett.\/} {\bf 95} 226801

\bibitem{Tang:2011by}
Tang E, Mei J~W and Wen X~G 2011 {\em Phys. Rev. Lett.\/} {\bf 106} 236802

\bibitem{Neupert:2011db}
Neupert T, Santos L, Chamon C and Mudry C 2011 {\em Phys. Rev. Lett.\/} {\bf
  106} 236804

\bibitem{Sun:2011dk}
Sun K, Gu Z~C, Katsura H and Das~Sarma S 2011 {\em Phys. Rev. Lett.\/} {\bf
  106} 236803

\bibitem{Regnault:2011bu}
Regnault N and Bernevig B~A 2011 {\em Phys. Rev. X\/} {\bf 1} 021014

\bibitem{Parameswaran:2012cu}
Parameswaran S~A, Roy R and Sondhi S~L 2012 {\em Phys. Rev. B\/} {\bf 85}
  241308

\bibitem{Goerbig:2012cz}
Goerbig M~O 2012 {\em Eur. Phys. J. B\/} {\bf 85} 15

\bibitem{2014PhRvB..90p5139R}
Roy R 2014 {\em Phys. Rev. B\/} {\bf 90} 165139

\bibitem{Jackson:2015fv}
Jackson T~S, M{\"o}ller G and Roy R 2015 {\em Nat Comm\/} {\bf 6} 8629

\bibitem{Spanton:2017vf}
Spanton E~M, Zibrov A~A, Zhou H, Taniguchi T, Watanabe K, Zaletel M~P and Young
  A~F 2017 {\em arXiv\/}  arXiv:1706.06116 (\textit{Preprint}
  \eprint{1706.06116})

\bibitem{Aidelsburger:2011hl}
Aidelsburger M, Atala M, Nascimb{\`e}ne S, Trotzky S, Chen Y~A and Bloch I 2011
  {\em Phys. Rev. Lett.\/} {\bf 107} 255301

\bibitem{2010PhRvA..82f3625M}
M{\"o}ller G and Cooper N~R 2010 {\em Phys. Rev. A\/} {\bf 82} 063625

\bibitem{Aidelsburger:2013ew}
Aidelsburger M, Atala M, Lohse M, Barreiro J~T, Paredes B and Bloch I 2013 {\em
  Phys. Rev. Lett.\/} {\bf 111} 185301

\bibitem{2013PhRvL.111r5302M}
Miyake H, Siviloglou G~A, Kennedy C~J, Burton W~C and Ketterle W 2013 {\em
  Phys. Rev. Lett.\/} {\bf 111} 185302

\bibitem{Chin:2013kc}
Chin C and Mueller E~J 2013 {\em Physics\/} {\bf 6} 118

\bibitem{Cooper:2018tr}
Cooper N~R, Dalibard J and Spielman I~B 2018 {\em arXiv\/} (\textit{Preprint}
  \eprint{1803.00249})

\bibitem{Jotzu:2014kz}
Jotzu G, Messer M, Desbuquois R, Lebrat M, Uehlinger T, Greif D and Esslinger T
  2014 {\em Nature\/} {\bf 515} 237--240

\bibitem{Duca:2015cz}
Duca L, Li T, Reitter M, Bloch I, Schleier-Smith M and Schneider U 2015 {\em
  Science\/} {\bf 347} 288--292

\bibitem{Aidelsburger:2015hm}
Aidelsburger M, Lohse M, Schweizer C, Atala M, Barreiro J~T, Nascimb{\`e}ne S,
  Cooper N~R, Bloch I and Goldman N 2015 {\em Nat Phys\/} {\bf 11} 162--166

\bibitem{Cooper:2011dq}
Cooper N~R and Dalibard J 2011 {\em European Physics Letters\/} {\bf 95} 66004

\bibitem{Sterdyniak:2015jo}
Sterdyniak A, Cooper N~R and Regnault N 2015 {\em Phys. Rev. Lett.\/} {\bf 115}
  116802

\bibitem{Sterdyniak:2015kp}
Sterdyniak A, Bernevig B~A, Cooper N~R and Regnault N 2015 {\em Phys. Rev. B\/}
  {\bf 91} 035115

\bibitem{Wu:2016kv}
Wu Z, Zhang L, Sun W, Xu X~T, Wang B~Z, Ji S~C, Deng Y, Chen S, Liu X~J and Pan
  J~W 2016 {\em Science\/} {\bf 354} 83--88

\bibitem{Huang:2016kf}
Huang L, Meng Z, Wang P, Peng P, Zhang S~L, Chen L, Li D, Zhou Q and Zhang J
  2016 {\em Nat Phys\/} {\bf 12} 540--544

\bibitem{Sun:2017wa}
Sun W, Wang B~Z, Xu X~T, Yi C~R, Zhang L, Wu Z, Deng Y, Liu X~J, Chen S and Pan
  J~W 2017 {\em arXiv\/} (\textit{Preprint} \eprint{1710.00717})

\bibitem{Windpassinger:2013is}
Windpassinger P and Sengstock K 2013 {\em Rep. Prog. Phys.\/} {\bf 76} 086401

\bibitem{Jo:2012br}
Jo G~B, Guzman J, Thomas C, Hosur P, Vishwanath A and Stamper-Kurn D 2012 {\em
  Phys. Rev. Lett.\/} {\bf 108} 045305

\bibitem{Hofstadter:1976wt}
Hofstadter D 1976 {\em Phys. Rev. B\/} {\bf 14} 2239--2249

\bibitem{Claro:1979tn}
Claro F and Wannier G~H 1979 {\em Phys. Rev. B\/} {\bf 19} 6068--6074

\bibitem{Rammal:1985ef}
Rammal R 1985 {\em J. Phys. France\/} {\bf 46} 1345--1354

\bibitem{Tarruell:2012db}
Tarruell L, Greif D, Uehlinger T, Jotzu G and Esslinger T 2012 {\em Nature\/}
  {\bf 483} 302--305

\bibitem{Uehlinger:2013dh}
Uehlinger T, Greif D, Jotzu G, Tarruell L, Esslinger T, Wang L and Troyer M
  2013 {\em Eur. Phys. J. Spec. Top.\/} {\bf 217} 121--133

\bibitem{Jaksch:1998ee}
Jaksch D, Bruder C, Cirac J, Gardiner C and Zoller P 1998 {\em Phys. Rev.
  Lett.\/} {\bf 81} 3108--3111

\bibitem{Vidal:1998gx}
Vidal J, Mosseri R and Doucot B 1998 {\em Phys. Rev. Lett.\/} {\bf 81}
  5888--5891

\bibitem{Korshunov:2005iq}
Korshunov S~E 2005 {\em Phys. Rev. Lett.\/} {\bf 94} 087001

\bibitem{2012PhRvL.108d5306M}
M{\"o}ller G and Cooper N~R 2012 {\em Phys. Rev. Lett.\/} {\bf 108} 045306

\bibitem{Payrits:2014ft}
Payrits M and Barnett R 2014 {\em Phys. Rev. A\/} {\bf 90} 013608

\bibitem{Andrijauskas:2015em}
Andrijauskas T, Anisimovas E, Ra{\v c}i{\=u}nas M, Mekys A, Kudria{\v s}ov V,
  Spielman I~B and Juzeliunas G 2015 {\em Phys. Rev. A\/} {\bf 92} 033617

\bibitem{Huber:2010bc}
Huber S~D and Altman E 2010 {\em Phys. Rev. B\/} {\bf 82} 184502

\bibitem{Mielke:1999jk}
Mielke A 1999 {\em Journal of Physics A: Mathematical and General\/} {\bf 32}
  8411

\bibitem{Noda:2014ku}
Noda K, Inaba K and Yamashita M 2014 {\em Phys. Rev. A\/} {\bf 90} 043624

\bibitem{Cooper:2012bt}
Cooper N~R and Moessner R 2012 {\em Phys. Rev. Lett.\/} {\bf 109} 215302

\bibitem{Fukui:2005ii}
Fukui T, Hatsugai Y and Suzuki H 2005 {\em Journal of the Physical Society of
  Japan\/} {\bf 74} 1674--1677

\bibitem{Ruostekoski:2002fs}
Ruostekoski J, Dunne G and Javanainen J 2002 {\em Phys. Rev. Lett.\/} {\bf 88}
  180401

\end{thebibliography}

\end{document}